\documentclass[times,twocolumn]{aastex631} 

\newcommand{\Msun}{\mbox{\,$\rm{M}_\odot$}}

\newcommand{\ltsimeq}{\raisebox{-0.6ex}{$\,\stackrel
	{\raisebox{-.2ex}{$\textstyle <$}}{\sim}\,$}}
\newcommand{\gtsimeq}{\raisebox{-0.6ex}{$\,\stackrel
	{\raisebox{-.2ex}{$\textstyle >$}}{\sim}\,$}}

\newcommand{\EBminusV}{\mbox{$E_{B-V}$}}

\newcommand{\fion}[2]{[{#1}\,{\sc {#2}}]}

\usepackage{color,soul}

\usepackage{graphics,graphicx}
\usepackage{epsfig}

\usepackage{datetime2}
\usepackage{amsmath}
\shorttitle{V1716 Sco IR Coronal Abundances}
\shortauthors{Woodward et al.}
\begin{document}
\title{A Photoionization model for the Infrared Coronal Line Emission in the Classical Nova V1716 Scorpii} 

\author[0000-0001-6567-627X]{C. E. Woodward$^{\dagger}$}
\affiliation{Minnesota Institute for Astrophysics, University of Minnesota,
116 Church Street SE, Minneapolis, MN 55455, USA}

\author[0000-0003-4615-8009]{G. Shaw}
\affiliation{Department of Astronomy and Astrophysics, Tata Institute of Fundamental Research, 
Homi Bhabha Road, Mumbai 4000005, India}

\author[0000-0002-1359-6312]{S. Starrfield}
\affiliation{School of Earth \& Space Exploration, Arizona State University, 
Box 876004, Tempe, AZ 85287-6004, USA}

\author[0000-0002-3142-8953]{A. Evans}
\affiliation{Astrophysics Group, Keele University, Keele, Staffordshire, ST5 5BG, UK}

\author[0000-0001-7796-1756]{K. L. Page}
\affiliation{School of Physics \& Astronomy, University of Leicester, Leicester, LE1 7RH, UK} 

\correspondingauthor{C.E. Woodward}
\email{mailto:chickw024@gmail.com}
\received{2023 Nov 09}
\accepted{2024 Apr 17}
\published{To appear in the Astrophysical J.}

\begin{abstract}
A near-infrared spectrum of nova V1716 Scorpii (PNV J17224490-4137160), 
a recent bright (V$_{max} = 7.3$ mag), Fermi-LAT detected $\gamma$-ray source, 
was modeled using the photoionization code CLOUDY. Abundances were estimated 
for He, C, N, O, Si, Al, Mg, Fe, Ne, S, Ca, and P. Notably, P (a factor of 120) and N 
(a factor of 248) are highly overabundant. It was necessary to 
assume the ejecta consist of two components (with a cylindrical geometry): a dense 
component from which the bulk of the H, He, and neutral O~I and N emission arises 
and a more diffuse component from which most of the coronal lines arise. Some of the 
coronal lines are found to originate from both the dense and diffuse components. The mass 
of the ejecta, including neutral and ionized gas, is $\simeq 4.19 \times 10^{-4}$~\Msun.
Our analysis indicates that in the case of V1716 Sco (which has a carbon-oxygen white dwarf), 
a fraction of 25\% white dwarf matter rather than 50\% is favored for the mixing between white 
dwarf and the accreted envelope before the outburst.  This mixing ratio is like that found for 
Oxygen-Neon novae where a 25\% mixing fraction is also indicated. Helium hydride -- the 
first molecule to form after the Big Bang -- may have formed in the ejecta of V1716~Sco based 
on photoionization modeling. This prediction suggests that novae may be potential formation 
sites of this important molecular ion.
\end{abstract}


\footnote{ $\dagger$ Visiting Astronomer at the Infrared Telescope Facility, which is operated 
by the University of Hawaii under contract 80HQTR19D0030 with the National Aeronautics and 
Space Administration.}

\section{Introduction}
\label{sec:sec-intro}
Classical nova (CN) explosions occur in semi-detached binary systems consisting of a white 
dwarf (WD) component and a main sequence dwarf. The latter fills its Roche Lobe, and 
consequently material spills on to the surface of the WD via an accretion disk. In time the 
material at the base of the accreted layer becomes degenerate, and hot enough to trigger 
a thermonuclear runaway (TNR). This results in a CN explosion, and the ejection of 
$\simeq 10^{-5}$  to $10^{-4}$ M$_\odot$ of material, enriched in C, N, O, Mg, Si, Al, Ne, 
and other metals, at several 100 to $\gtsimeq 1000$ km s$^{-1}$ \citep[e.g.,][]{2012clno.book.....B, 2012BASI...40..161A}.
In view of the Galactic CN rate \citep[$\simeq 47$ yr$^{-1}$; ][]{2021ApJ...912...19D}, likely CNe 
are a major source of $^{13}$C, $^{15}$N and $^{17}$O in the Galaxy \citep{1998PASP..110....3G, 2012clno.book.....B} 
and may make a significant contribution to Galactic $^{7}$Li \citep{2020ApJ...895...70S, 2024ApJ...962..191S}, although 
the case for this rests on observational results and not 
theory \citep{2020A&A...634A...5J, 2022ApJ...933L..30K, 2022MNRAS.509.3258M,  2023MNRAS.518.2614M}.

As the ejecta disperse, an emission line spectrum is produced and stable nuclear burning 
causes the pseudophotosphere to shrink, revealing a hotter, deeper source of emission. 
CNe spectra are remarkable for their changing elemental and ion content and the 
temporal development of line profiles is critical to understanding the dynamics of ejection. 
Low-energy permitted lines of CNO and Fe II give way to He II, as well as high ionization lines, 
e.g., \fion{Fe}{VII} 6087~\AA, and ultimately to infrared (IR) ``coronal'' 
lines \citep{2015AJ....149..136R, 2021ApJ...922L..10W, 2022MNRAS.510.4265K}. The latter 
lines are sources of abundance information as a wide range of isoelectronic 
sequences \citep{1990ApJ...352..307G} and adjacent ionization states of metals are 
observable. Often, as the ejecta cool and evolve, molecules 
\citep[e.g., CO --][]{2003ApJ...596.1229R, 2004MNRAS.347.1294P, 2009MNRAS.398..375D, 2016MNRAS.455L.109B}
and dust form. CNe originating on CO WDs are often dust-formers and, while C is a major 
grain component, silicates, polycyclic aromatic hydrocarbons (PAHs), and SiC are 
often present, occasionally in the same nova \citep{2012BASI...40..213E}. 

The evolution of the TNR depends upon the mass and luminosity of the WD, the rate of mass 
accretion, the composition of the accreted material, and the chemical composition in the reacting layers.
Hydrodynamic studies of the accretion process on to the WD preceding the TNR event and 
the degree to which material from the underlying WD and the envelope is admixed in the ejecta
has been studied by many groups for several decades 
\citep{1995ApJ...448..807P, 2012MNRAS.427.2411G, 2013ApJ...777..130K, 2020A&A...634A...5J, 2020ApJ...895...70S, 2024ApJ...962..191S}.
However, the constraints on the theoretical models of nucleosynthesis in the outburst, chemical 
anomalies related to nucleosynthesis, and the evolution of the progenitor are provided by spectroscopic
observations of the ejecta from which detailed elemental abundance patterns can be derived.

In this paper we estimate ejecta abundances for V1716 Sco in 
the coronal line phase of its evolution 132.8 days after outburst \citep{2023ATel16222....1W}  
derived from photoionization modeling of IR (0.7 to 4.2~\micron) spectra using 
CLOUDY \citep{1998PASP..110..761F, 2023RMxAA..59..327C}.  These 
abundance patterns are compared with theoretical simulations 
of CNe  \citep{2020ApJ...895...70S, 2024ApJ...962..191S} with differing core-envelope 
mixing ratios and their ejecta abundance patterns to understand the characteristics of the 
underlying WD. Since the coronal phase is fairly prevalent in novae \citep[e.g., the statistics 
compiled by ][]{1990AJ....100.1588B},  the present work may serve as a useful 
template for comparing results from similar near-IR modeling of CNe that may erupt 
in the future.

\section{V1716~Sco}
\subsection{General Properties}
\label{sec:sec-properties}

Nova V1716 Sco (PNV J17224490-4137160) was discovered on 2023 April 20.6780 UT by
A. Pearce.\footnote{\url{http://www.cbat.eps.harvard.edu/unconf/followups/J17224490-4137160.html}}
The nova was bright on 2023 April 20.410 UT \citep[MJD 60054.410;][]{2023ATel16018....1S}, which 
we take as our origin of time ($t_{o}$). Its CN status was confirmed by \citet{2023ATel16003....1W}, 
who described it as a lightly-reddened Fe II nova near maximum light. Optical spectroscopy 
of the earliest phases was obtained by \citet{2023a_ATel16004....1S, 2023b_ATel16006....1S, 2023c_ATel16036....1S} 
and \citet{2023ATel16007....1I}. The early optical spectra showed ejection velocities 
$\simeq 1700$~km~s$^{-1}$,  although there were emission components extending 
to $\pm 3000$~km~s$^{-1}$. The Neil Gehrels Swift Observatory \citep{2004ApJ...611.1005G}
has observed V1716~Sco regularly since outburst; details will be presented elsewhere.
NuSTAR observations of the hard X-ray spectrum of V1716~Sco early in the
outburst \citep{2023ATel16018....1S}  revealed a heavily absorbed thermal plasma. 
V1716 Sco joins the increasing inventory of CNe that are $\gamma$-ray 
sources \citep{2015ICRC...34..880C, 2018cosp...42E.625C}. 

V1716~Sco is a fast nova \citep[for speed-class definition, see][]{2021ARA&A..59..391C}. The AAVSO 
V-band light curve yields $t_{2}$ and $t_{3}$ values ($t_{2}$ and $t_{3}$ are the times to 
decline by 2 and 3 magnitudes respectively from peak brightness) of $\sim 5.8$ and 
11.7~d respectively.

\subsection{Distance and Reddening}
\label{sec:sec-dist-red}

The reddening, \EBminusV, to V1716 Sco was estimated by \citet{2023b_ATel16006....1S} to 
be in the range $0.45 \ltsimeq$ \EBminusV $\ltsimeq 0.6$, consistent with the lower
limit (\EBminusV $\gtsimeq 0.5$ from diffuse interstellar absorption features) given by \citet{2023ATel16007....1I}. 
The methods of \citet{1987A&AS...70..125V} using the intrinsic colors of CNe 
give \EBminusV \, = 0.66 (color at maximum light) and \EBminusV \, = 0.70 (color at time $t_{2}$). 
Using the simultaneous solution for A$_{\rm{V}}$ and distance, D, method described in \citet{2022MNRAS.510.4265K},
we obtain D $= 3.6 \pm 0.6$~kpc and a somewhat higher value of reddening, \EBminusV \, = 1.0. 
The relative strengths of the OI 0.8446 and 1.1287~\micron{} lines can be used to 
estimate the reddening to the nova \citep{2016MNRAS.462.2074S}. Their ratio, as 
obtained from the spectrum presented here, suggests  \EBminusV \, $\sim 0.57$
using a \citet{1989ApJ...345..245C} interstellar extinction law. For the 
present work it seems reasonable to adopt a value of \EBminusV \, = 0.65 
(the average of all estimates is $0.64 \pm 0.18$).

For a value of \EBminusV \, = 0.65, the distance, D, is found to be $\sim$ 2.1~kpc from the 
extinction versus distance relation derived by \citet{2006A&A...453..635M}.  A 19$^{\rm{th}}$ 
magnitude star (Gaia DR3 5959616875349110656) is found to closely match the position 
of V1716~Sco (the offset between this star and the nova is 0\farcs42). This star 
has a parallax of 1.0863~mas and parallax error of 0.3907~mas 
in the EDR3 database \citep{2023A&A...674A...1G}. A search of the Dark Energy 
Camera Plane Survey \citep[DECaPS,][]{2023ApJS..264...28S} using a 2\farcs0 cone search
around the position of V1716~Sco with the Astro Data Lab query interface \citep{2020A&C....3300411N} 
and the DECaPS~DR2 did not return any cataloged points sources. We assume that the Gaia 
positionally associated source is the progenitor  \pagebreak

\begin{figure*}[hbt!]
\figurenum{1}
\begin{center}
\includegraphics[trim=0.3cm 0.5cm 0.5cm 0.25cm, clip, width=0.98\textwidth]{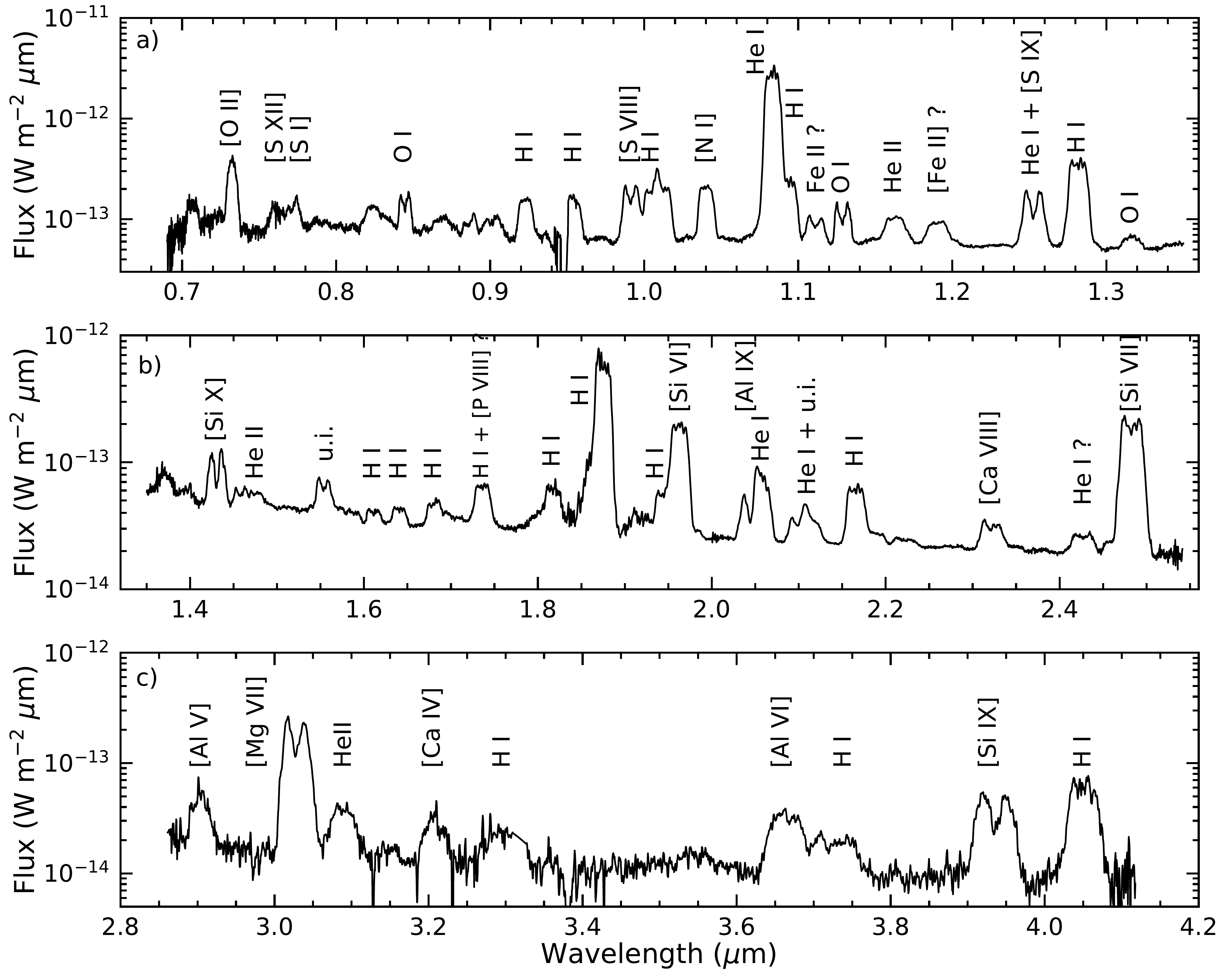}
\caption{The observed spectrum of V1716~Sco on day +132.8, with the prominent lines identified.
a)~the 0.66 to 1.34~\micron{} section (SpeX SXD). b)~the 1.35 to 2.5~\micron{} section (SpeX SXD), 
and c) the 2.80 to 4.20~\micron{} section. For these latter data, the observed spectral segment (SpeX LXD) was 
smoothed with a Savtisky-Golay filter of window width = 7, polynomial order = 2, to improve the 
SNR and each end of the spectral segments were clipped (where the SNR $\ltsimeq 3$).  
Unidentified lines are indicated with 'u.i.' The observed spectrum (DbF) for each panel is available
as a machine readable table (MRT) in the on-line manuscript.
}
\label{fig:fig-one} 
\end{center}
\end{figure*}

\noindent star of the nova. As recommended for the 
Gaia data, when the fractional parallax uncertainties are not too large (as applicable in this case) 
use of the Gaia geometric distance ($3.17^{+2.2}_{-1.6}$~kpc) is suggested rather than that of 
the Gaia photogeometric distance, which in this case is $4.97^{+1.7}_{-1.1}$~kpc \citep{2021AJ....161..147B}.  
Thus for the present work we adopt D $\simeq 3$~kpc (a value close to the Gaia geometric distance),
although there is considerable uncertainty.

\section{Observations}
\label{sec-obs}
A 0.7 to 4.2~\micron{} spectrum of V1716 Sco was obtained on 2023 August 31.24 UT 
(MJD 60187.24) on the 3.2 m NASA Infrared Telescope Facility (IRTF). Observations were 
obtained with the medium-resolution facility spectrograph 
\citep[SpeX,][]{2003PASP..115..362R} in the cross-dispersed short (SXD) and long (LXD\_\,short) 
modes to cover the spectral range of 0.7 - 4.2~\micron. The observations were made with a 
0\farcs5 slit ($R = 1200$), at an airmass between 2.10 – 2.24, under photometric conditions 
and moderate seeing ($\ltsimeq$ 1\farcs1 in the K-band). The A0V star HD~148418 was
used to correct for telluric absorption and the total on-source integration times for V1716~Sco 
were 1978s and 556s respectively (the SXD and LXD modes). The SpeX data were reduced and 
calibrated using Spextool \citep{2004PASP..116..362C} and the tool xtellcor \citep{2003PASP..115..389V} 
was used for the corrections for telluric absorption. The observed spectrum is presented in 
Figure~\ref{fig:fig-one}.

\begin{figure*}[htb!]
\figurenum{2}
\begin{center}
\includegraphics[angle=-90, trim=1.5cm 0.25cm 1.0cm 2.5cm, clip, width=0.58\textwidth]{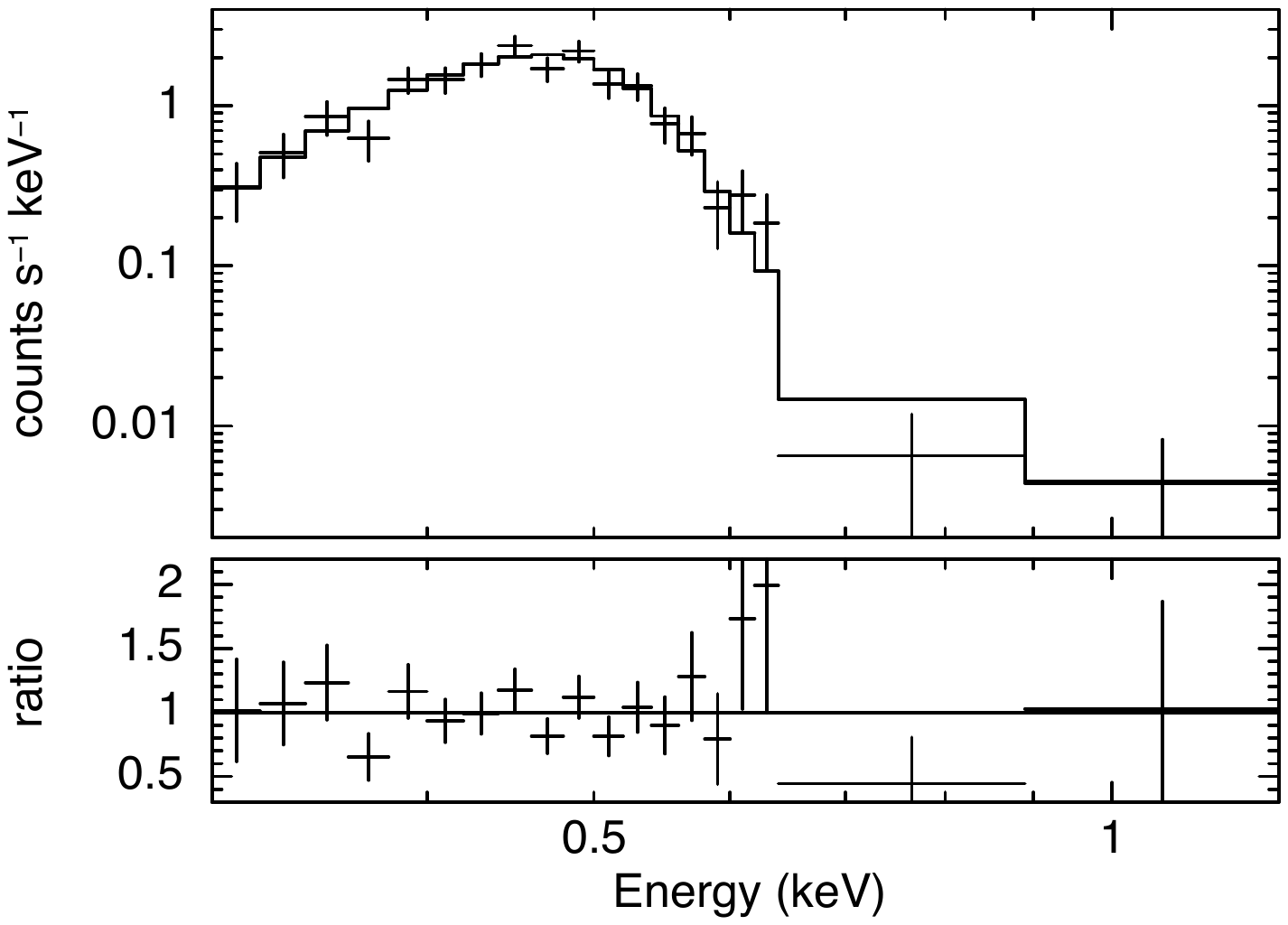}
\caption{The Swift x-ray spectrum of V1716~Sco obtained contemporaneously with 
the near-IR spectrum. The top panel shows a blackbody fit using HEASOFT XSPEC tools
\citep{1996ASPC..101...17A} to the soft emission with 
kT$ = 23.6^{+3.1}_{-3.6}$ eV (or log(T) = 5.44~K). The bottom panel is the ratio of the data to the model fit. 
}
\label{fig:fig-two} 
\end{center}
\end{figure*}

The Swift soft X-ray spectrum of 31 August 2023 (MJD 60187.33083 $\pm$ 0.00627; obtained within
2~hrs of the IR spectrum) was  parametrized by a blackbody and provided an initial estimate 
for T$_{\rm{BB}}$(K). The spectrum is shown in Figure~\ref{fig:fig-two} fitted by a BB 
with kT = $23.6^{+3.1}_{-3.6}$ eV, or log~T~(K)  = 5.44.  As an initial estimate for the bolometric 
luminosity $L_{\rm{bol}}$, the BVRI data at maximum from AAVSO were dereddened and 
fitted by a blackbody to give a temperature of $\sim 8400$~K equivalent to an outburst 
luminosity  $\simeq 10^{38.45}$~ergs~s$^{-1}$ for a distance of 3~kpc.  The latter BB has a radius
of $8.9\times10^{12}$~cm (adopting an expansion velocity of 1000~km~s$^{-1}$ and $t = 130$d). 

\begin{figure*}[htb!]
\figurenum{3}
\begin{center}
\includegraphics[angle=0, trim=0.5cm 0.05cm 0.5cm 0.2cm, clip, width=0.58\textwidth]{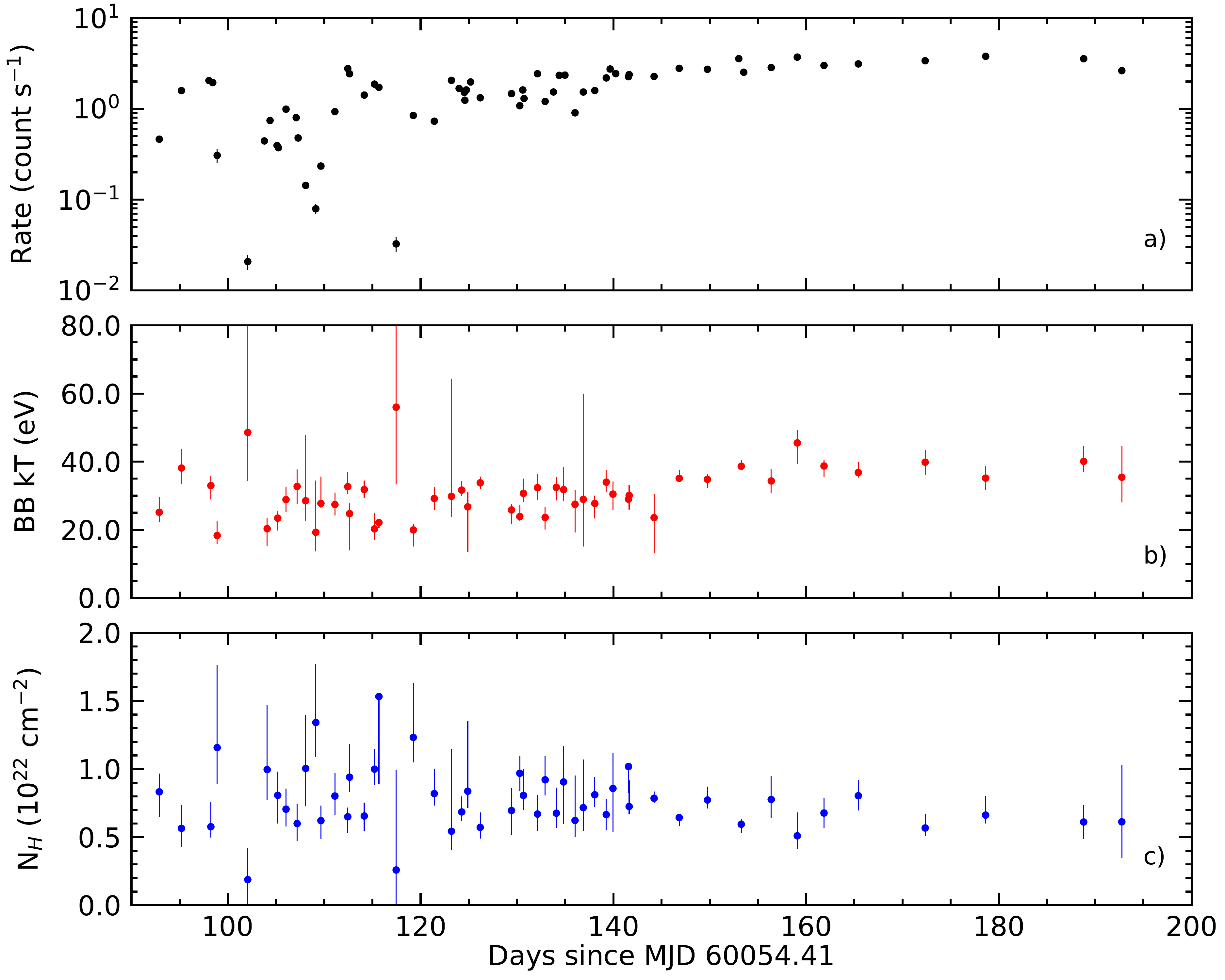}
\caption{Synoptic Swift observations of V1716~Sco post outburst and spectral
fits. a)~The light curve. b)~The derived kT (eV) from model fits assuming a simple blackbody
for the SSS for the soft x-ray spectra (red points with associated asymmetric error). Only a 
modest variance in the overall temperature of the x-ray source from day 133 through 
day 190 is evident. c)~The derived hydrogen column density (blue dots 
with associated asymmetric errors). The data for each panel are available as a machine 
readable table (MRT) in the on-line manuscript.} 
\label{fig:fig-xrt-bb-fits} 
\end{center}
\end{figure*}

V1716~Sco entered a super-soft x-ray (SSS) phase near day 55 \citep{2023ATel16069....1P}.
Spectral fits (assuming a simple BB for the SSS source) to all the Swift x-ray spectra 
suggest only a slight increase in the temperature ($\simeq 30$~eV)
between day 133 and 180 but nothing significant, as shown in Figure~\ref{fig:fig-xrt-bb-fits}.

\begin{deluxetable}{@{\extracolsep{0pt}}lr}
\tablenum{1}
\setlength{\tabcolsep}{3pt} 
\tablecaption{V1716 Sco CLOUDY Parameters\label{tab:table-1}}
\tablehead{
\colhead{Cylindrical } & \colhead{Day}\\
\colhead{Model} & \colhead{+132.8}
}
\startdata
log T$_{BB}$(K)& 5.74 \\
log L(erg s$^{-1}$) & 37.5\\
log H$_{\rm{den}}$(LD - diffuse)(cm$^{-3}$) & 6.40 \\ 
log H$_{\rm{den}}$(HD - clump)(cm$^{-3}$) & 7.85 \\
log R$_{\rm{in}}$(cm)& 15.01\\
log R$_{\rm{out}}$(cm)& 15.3367\\
log h (cm) & 15.35\\
Covering factor & 0.5\\
Filling factor (diffuse)& 0.80\\
Filling factor (clump)& 0.20 \\
\sidehead{\underbar{Abundances}\,\tablenotemark{a)} }
He/H & 18.72 $\times$ 10$^{-2}$\\
C/H & 17.41 $\times$ 10$^{-4}$\\
N/H & 167.65 $\times $10$^{-4}$\\
O/H & 93.09 $\times$ 10$^{-4}$ \\
Ne/H & 2.18 $\times$ 10$^{-4}$\\
S/H & 98.99 $\times$ 10$^{-6}$\\
Si/H & 71.28 $\times$ 10$^{-6}$ \\
Al/H & 16.92 $\times$ 10$^{-6}$\\
Mg/H & 11.94 $\times$ 10$^{-5}$\\
Ca/H & 5.26 $\times$ 10$^{-6}$\\
P/H & 30.84 $\times$ 10$^{-6}$\\ 
Fe/H & 63.20 $\times$ 10$^{-6}$\\
\\
Total number of lines & 39\\
Number of free parameters & 16\\
Degrees of freedom & 23\\
Reduced $\chi$ $^2$ & 2\\
\enddata
\tablenotetext{a)\, }{\, Cloudy23.01 model abundances are optimized assuming the abundances given in \citet{2010ApSS.328..179G}.}
\tablecomments{All other elements have non-depleted (with respect to H) solar abundances.}
\end{deluxetable}

\section{Results}
\label{sec:sec-res}

\subsection{Photoionization Models}
\label{sec:sec-cloudy}

\subsubsection{Photoionization Models - Cylindrical Geometry} 

The photoionization spectral synthesis code CLOUDY \citep[version C23.01][]{2023RMxAA..59..327C}
was used to model the IR spectroscopic data of V1716~Sco. We assume the surface of 
the central WD is emitting ionizing blackbody radiation with a temperature T$_{BB}$(K) 
and a bolometric luminosity, $L_{\rm{bol}}$ (erg~s$^{-1}$),  irradiating a cylindrical geometry 
of gas expanding with a velocity of v$_{\rm{exp}}$ (km~s$^{-1}$).  The cylindrical model in 
CLOUDY is basically a truncated spherical model.The dimension of this gas 
is defined by an inner radius r$_{\rm{in}}$ (cm), thickness $r_{\rm{d}}~{\rm (cm)}$, 
and the height of the cylinder $h~{\rm (cm)}$.  CNe often exhibit asymmetric geometries 
that are highly non-spherical, containing knots and clumps of material 
\citep[for example V1280~Sco, T Pydixis, or RR Pic,][]{2012A&A...545A..63C, 2013ApJ...768...48T, 2024A&A...681A.106C}
motivating use of a cylindrical model. The solar abundances called in the Cloudy modeling are from
\citet{2010ApSS.328..179G}. 

In the model presented here, we assume a number density law proportional to
n$_{\rm{in}}$r$^{-3}$, where n$_{\rm{in}}$(cm$^{-3}$) is the total hydrogen number density,

\begin{equation}
\rm{n}_{\rm{in}}(\rm{H}^{0}) + n_{\rm{in}}(\rm{H}^{+}) + 2 \times \rm{n}_{\rm{in}}(\rm{H}_{2}) + 
\Sigma_{\rm{other}}\, n_{\rm{in}}(H_{\rm{other}}),
\end{equation}

\noindent the latter term being the
summation of other species containing hydrogen nuclei such as H$_{3}^+$, H$_{2}^+$, etc.
The observed  v$_{\rm{exp}}$ ranges over 1000 to 1350~km~s$^{-1}$. We adopt
v$_{\rm{exp}} = 1000$~km~s$^{-1}$. Hence, both r$_{\rm{in}}$ (cm) and $r_{\rm{d}}~{\rm (cm)}$ 
are set by observation. However, the cylindrical height is a free parameter.  The 
physical parameters and their values for our final ``best-fit'' model are listed in Table~\ref{tab:table-1}.
Later, we consider the observed velocity range as well as an additional
model variable.

The observed spectrum exhibits highly ionized lines (Si X, Si IX) and neutral lines (H I, O I, He I). 
The physical processes of forming these lines differ.  In the default mode, CLOUDY considers 
both photoionization and collisional ionization. The collisional ionization rate coefficients are 
from \citet{1997ADNDT..65....1V} and \citet{2007A&A...466..771D}.  In the current model 
both photoionization and collisional ionization processes are enabled, so strictly it is not a 
pure photoionization model. However, most of the lines are generated through photoionization 
and recombination. CLOUDY has options to use these processes separately. However, for V1716~Sco,
the combined photoionization and collisional model works better (as compared to the observations) than 
a solely collisional or solely photoionization model. 

We assume a two component model, consisting of high (``clump'') and low densities (``diffuse''). 
The highly ionized lines arise from the low density component (A), whereas the neutral lines arise from the high 
density component (B). While the density and the filling factors for these two components are different, 
the other modeling parameters are the same.  Our final model predicts an average hydrogen density
at the inner radius, r$_{\rm{in}}$ for components A and B to be log~H$_{\rm{den}}$ (cm$^{-3}$) = 6.4 
and 7.85, respectively. Component A and component B contain 12$\%$ and 88$\%$ of the volume
of the ejecta, respectively. In component B, a density higher than log~H$_{\rm{den}}$ = 7.85 reduces 
the forbidden line fluxes due to increased collisional de-excitations.

We find that a blackbody of temperature 10$^{5.74}$ K with bolometric luminosity 
10$^{37.5}$ erg s$^{-1}$ is required to reproduce the observed line intensities on day 133. 
This high blackbody temperature is necessary to reproduce highly ionized line fluxes, such as 
for Si IX. On the other hand, a luminosity of 10$^{37.5}$ erg s$^{-1}$ makes the forbidden lines 
much weaker and the H I lines much stronger than is observed. The physical parameters 
and their values for the cylindrical model best-fit are listed in Table~\ref{tab:table-1}. We 
vary only the elemental abundances of the observed lines (He, O, N, Mg, P, Ca, Al, S, Si). 
Most of the predicted line intensities match with their observed values within the 
observed uncertainties (see Table~\ref{tab:table-2-cylindar}).

\subsubsection{Photoionization Models - Spherical Geometry} 

\begin{figure}[htb!]
\figurenum{4}
\begin{center}
\includegraphics[trim=0.015cm 0.95cm 0.15cm 0.15cm, clip, width=0.45\textwidth]{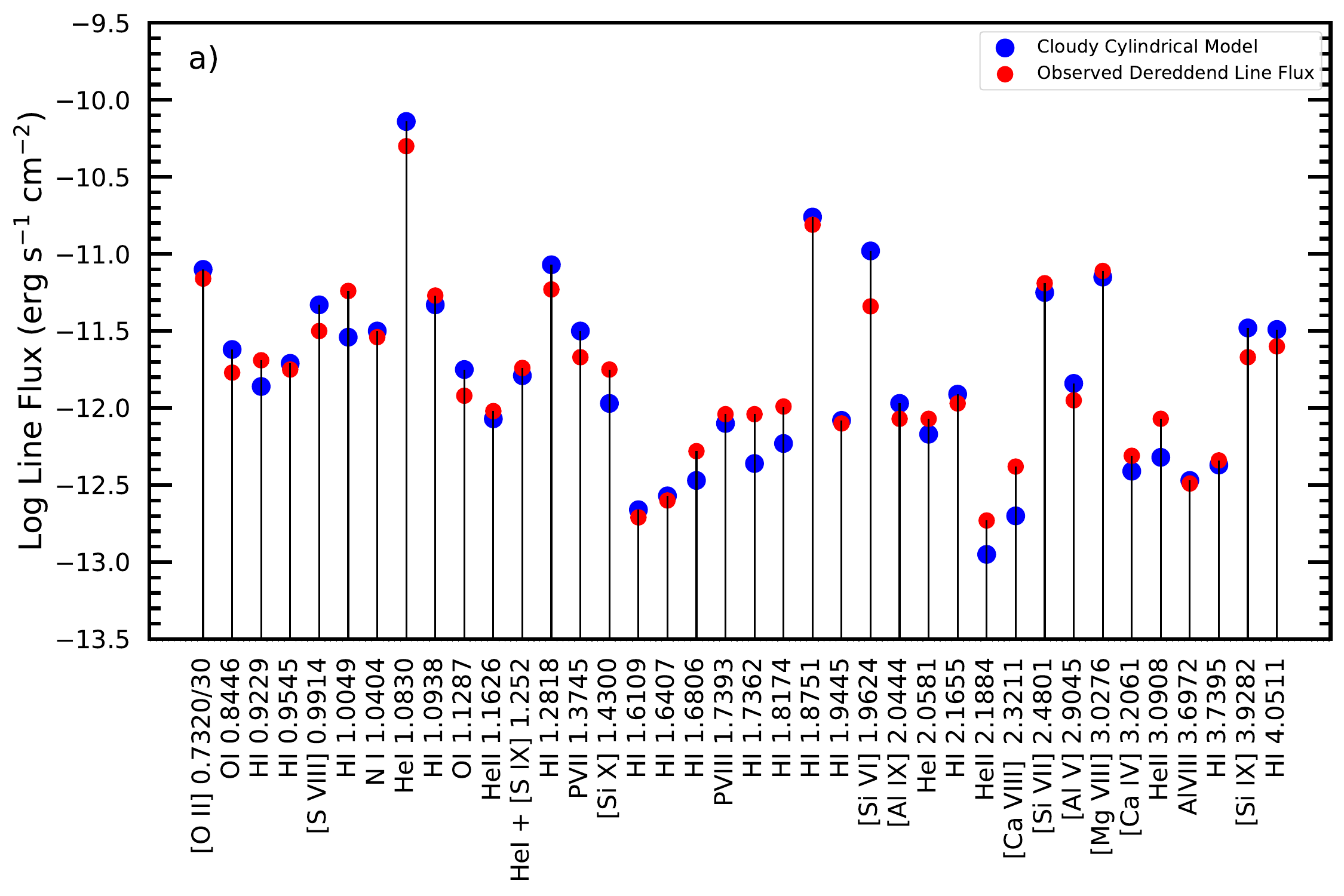}
\includegraphics[trim=0.015cm 0.95cm 0.15cm 0.15cm, clip, width=0.45\textwidth]{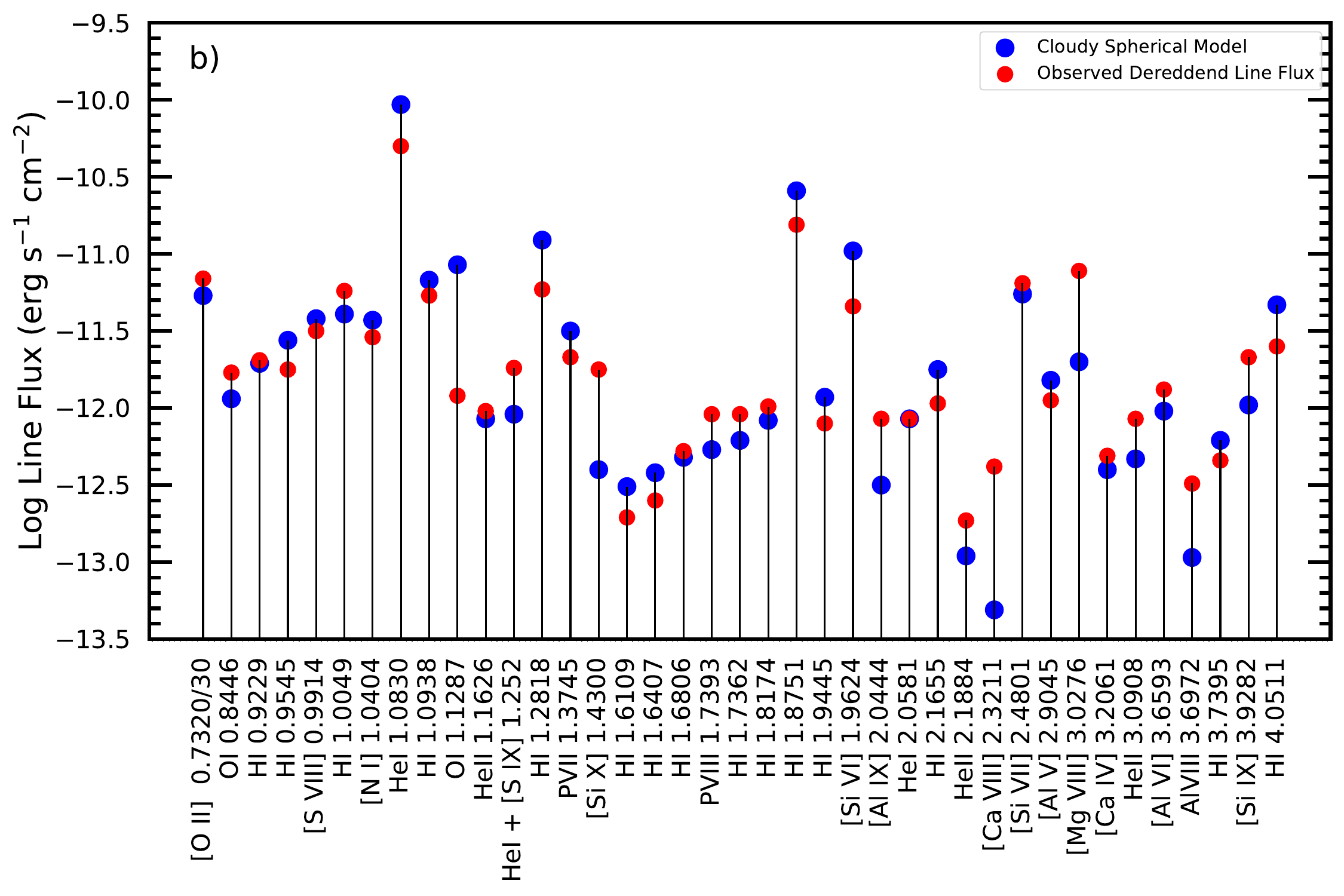}
\includegraphics[trim=0.015cm 0.95cm 0.15cm 0.15cm, clip, width=0.45\textwidth]{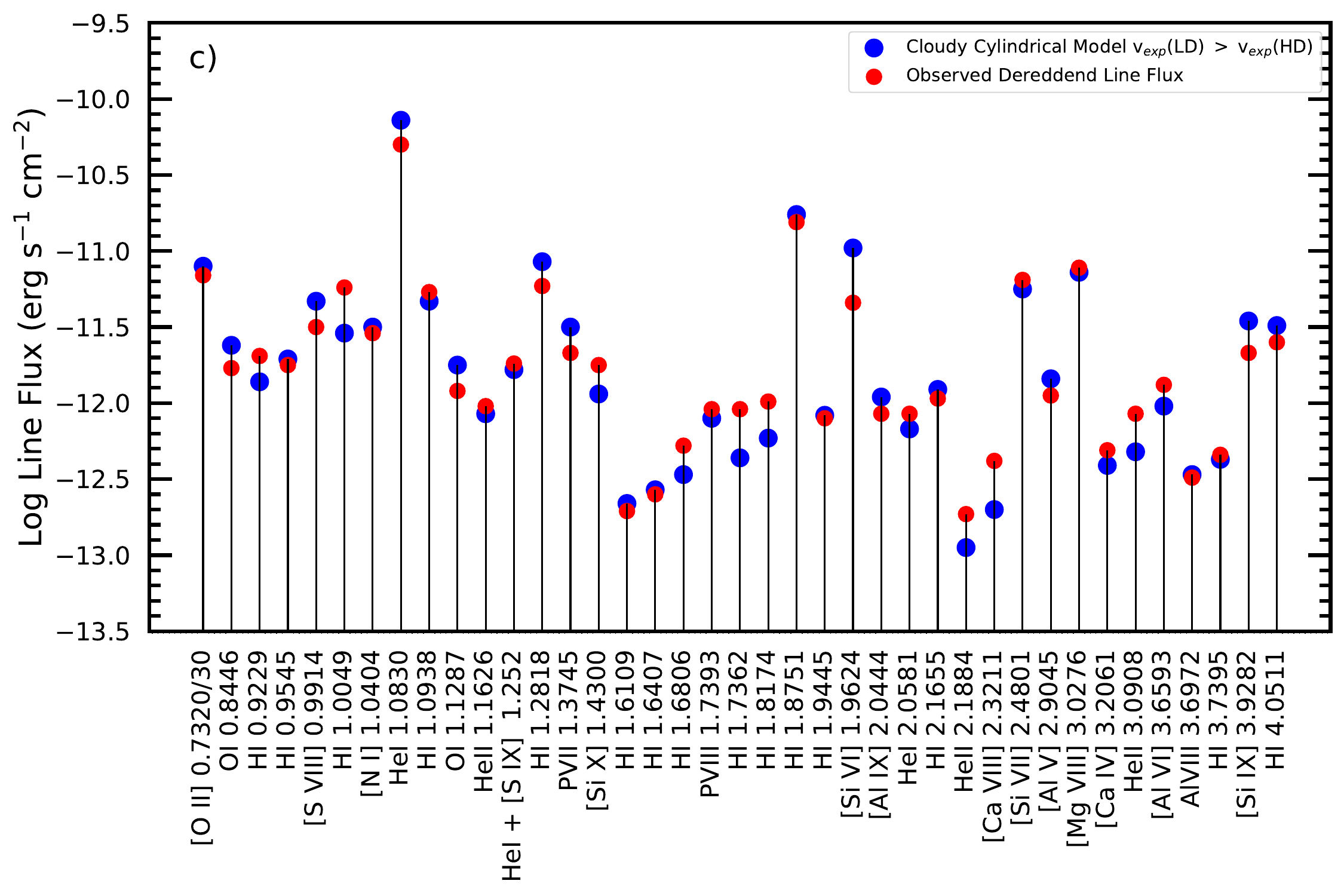}
\caption{Observed dereddened line fluxes (red filled circles) versus CLOUDY line fluxes (blue filled circles).
a) From the best-fit cylindrical model (Table~\ref{tab:table-2-cylindar}).  b) From a spherical model (Table~\ref{tab:table-3-sphere}).
c) From the best-fit cylindrical model (Table~\ref{tab:table-4-ldhd}) where the LD component is expanding with
a velocity greater than the HD component.
}
\label{fig:fig-threeA} 
\end{center}
\end{figure}

As a check, we consider a spherical geometry of the ejecta. In Table~\ref{tab:table-3-sphere}, 
we present our predictions for the spherical model keeping all the input parameters the same 
as the cylindrical model. Our model predictions with a cylindrical geometry match better 
with the observations despite the cylindrical model being a very simple truncated 
spherical model (see Fig.~\ref{fig:fig-threeA}). Hence, we consider the cylindrical 
model as our best fit model. 

%

\subsubsection{Photoionization Models - Ejecta Velocity} 
Observation reveals that v$_{\rm{exp}}$ ranges from 1000 -1350~km~s$^{-1}$,  
i.e., close to half the FWHM of the hydrogen Paschen-$\beta$ line of 2700~km~s$^{-1}$. 
Generally the v$_{\rm{exp}}$ is typically half the FWHM for any expanding
spherical shell of gas \citep{1982MNRAS.199..649R}. Hence, for 
the cylindrical model, we assume the low density component expanding at 1350~km~s$^{-1}$ 
which is greater than the expansion velocity of the dense component. The results are 
listed in Table~\ref{tab:table-4-ldhd}. This model increases the \fion{Si}{X} line flux by 0.05 dex.  
Considering all  the three cases mentioned above, (see Fig.~\ref{fig:fig-threeA}), 
a cylindrical model with low density region expanding with a higher velocity than the high
 density region seems to be the best fit model. 

%
\begin{deluxetable*}{@{\extracolsep{0pt}}lcccc}
\tablenum{2}
\setlength{\tabcolsep}{3pt} 
\tablecaption{V1716 Sco Observed and CLOUDY Line Luminosities for the Cylindrical Model  \label{tab:table-2-cylindar}}
\tablehead{
\colhead{} & \colhead{log flux\tablenotemark{$\dagger$} } & \colhead{log flux\tablenotemark{$\dagger$}} & \colhead{log total flux}  & \colhead{Dereddend} \\  
\colhead{Line} & \colhead{(LD)} & \colhead{(HD)} & \colhead{(HD + LD)} & \colhead{Observed Flux}\\
\colhead{($\mu$m)} & \colhead{(erg s$^{-1}$ cm$^{-2}$)} & \colhead{(erg s$^{-1}$ cm$^{-2}$)} & \colhead{(erg s$^{-1}$ cm$^{-2}$)} & \colhead{(erg s$^{-1}$ cm$^{-2}$)}
}
\startdata
\fion{O}{II} 0.7320/30	&	-20.13	&	-11.10	&	-11.10 &	-11.16	\\
OI 0.8446	&	$<$-20.00	&	-11.62	&	-11.62&	-11.77	\\
HI 0.9229	&	-13.96	&	-11.86	&	-11.86&	-11.69	\\
HI 0.9545	&	-13.80	&	-11.72	&	-11.71&	-11.75	\\
\fion{S }{VIII]} 0.9914	&	-11.91	&	-11.47	&	-11.33  &	-11.50	\\
HI 1.0049	&	-13.62	&	-11.54	&	-11.54	&	-11.24	\\
\fion{N}{I}1.0404	&	$<$-20.00	&	-11.50	&	-11.50&	-11.54	\\
HeI 1.0830	&	-14.99	&	-10.14	&	-10.14 &	-10.30	\\
HI 1.0938	&	-13.42	&	-11.33	&	-11.33 &	-11.27	\\
OI 1.1287	&	$<$-20.00	&	-11.75	&	-11.75	&	-11.92	\\
HeII 1.1626	&	-13.39	&	-12.09 &	-12.07	&	-12.02	\\
HeI  + \fion{S}{IX}  1.252	&	-11.93	&	-12.36	&	-11.79	&	-11.74	\\
HI 1.2818	&	-13.13 &	-11.07	&	-11.07	&	-11.23	\\
PVII 1.3745	&	-13.43	&	-11.50	&	-11.50	&	-11.67	\\
\fion{Si}{X} 1.4300	&	-11.99	&	-16.21	&	-11.97	&	-11.75	\\
HI 1.6109	&	-14.81	&	-12.66	&	 -12.66&	-12.71	\\
HI 1.6407	&	-14.70 &	-12.57	&	-12.57	&	-12.60	\\
HI 1.6806	&	-14.58	&	-12.47	&	-12.47	&	-12.28	\\
PVIII 1.7393	&	-12.47	&	-12.35	&	-12.10&	-12.04	\\
HI 1.7362	&	-14.47	&	-12.36	&	-12.36	&	-12.04	\\
HI 1.8174	&	-14.33	&	-12.23	&	-12.23	&	-11.99	\\
HI 1.8751	&	-12.86	&	-10.76	&	-10.76	&	-10.81	\\
HI 1.9445	&	-14.17	&	-12.08	&	-12.08&	-12.10	\\
\fion{Si}{VI} 1.9624	&	-14.50	&	-10.98 & -10.98 &	-11.34	\\
\fion{Al}{IX} 2.0444	&	-11.97	&	-14.48&	-11.97	&	-12.07	\\
HeI 2.0581	&	-19.04	&	-12.17	&	-12.17&	-12.07	\\
HI 2.1655	&	-13.99	&	-11.91	&	-11.91	&	-11.97	\\
HeII 2.1884	&	-14.26	&	-12.97	&	-12.95	&	-12.73	\\
\fion{Ca}{VIII}  2.3211	&	-12.71	&	-14.47 &	-12.70 &	-12.38	\\
\fion{Si}{VII} 2.4801	&	-12.59 &	-11.27	&	-11.25	&	-11.19	\\
\fion{Al}{V} 2.9045	&	-17.24	&	-11.84	&	 -11.84	&	-11.95	\\
\fion{Mg}{VIII} 3.0276	&	-11.16	&	-13.02 &	-11.15 &	-11.11	\\
\fion{Ca}{IV} 3.2061	&	-18.70	&	-12.41	&	-12.41	&	-12.31	\\
HeII 3.0908	&	-13.66	&	-12.34	&	-12.32 &	-12.07	\\
\fion{Al}{VI} 3.6593	&	-14.28	&	-12.02	&	-12.02	&	-11.88	\\
AlVIII 3.6972	&	-12.50 &	-13.72 &	-12.47 &	-12.49	\\
HI 3.7395	&	-14.46	&	-12.37 &	-12.37, &	-12.34	\\
\fion{Si}{IX}  3.9282	&	-11.48	&	-14.96 &	 -11.48 &	-11.67	\\
HI 4.0511	&	-13.55	&	-11.50	&	 -11.49 &	-11.60	\\
\enddata
\tablecomments{$^{\dagger}$ LD/HD = low/high density.}
\end{deluxetable*}

%
\begin{deluxetable*}{@{\extracolsep{0pt}}lcccc}
\tablenum{3}
\setlength{\tabcolsep}{3pt} 
\tablecaption{V1716 Sco Observed and CLOUDY Line Luminosities for the Spherical Model$^{\ddagger}$ \label{tab:table-3-sphere}}
\tablehead{
\colhead{} & \colhead{log flux\tablenotemark{$\dagger$} } & \colhead{log flux\tablenotemark{$\dagger$}} & \colhead{log total flux}  & \colhead{Dereddend} \\  
\colhead{Line} & \colhead{(LD)} & \colhead{(HD)} & \colhead{(HD + LD)} & \colhead{Observed Flux}\\
\colhead{($\mu$m)} & \colhead{(erg s$^{-1}$ cm$^{-2}$)} & \colhead{(erg s$^{-1}$ cm$^{-2}$)} & \colhead{(erg s$^{-1}$ cm$^{-2}$)} & \colhead{(erg s$^{-1}$ cm$^{-2}$)}
}
\startdata
\fion{O}{II} 0.7320/30	&	-20.60	&	-11.27	&	-11.27 &	-11.16	\\
OI 0.8446	&	$<$-20.00	&	-11.94	&	-11.94 &	-11.77	\\
HI 0.9229	&	-14.52	&	-11.71	&	-11.71&	-11.69	\\
HI 0.9545	&	-14.36	&	-11.56	&	-11.56 &	-11.75	\\
\fion{S}{VIII} 0.9914	&	-12.48	&	-11.46	&	-11.42  &	-11.50	\\
HI 1.0049	&	-14.18	&	-11.39	&	-11.39	&	-11.24	\\
\fion{N}{I} 1.0404	&	$<$-20.00	&	-11.43	&	-11.43&	-11.54	\\
HeI 1.0830	&	-15.37	&	-10.03	&	-10.03 &	-10.30	\\
HI 1.0938	&	-13.97	&	-11.17	&	-11.17 &	-11.27	\\
OI 1.1287	&	$<$-20.00	&	-11.07	&	-11.07	&	-11.92	\\
HeII 1.1626	&	-13.95	&	-12.09 &	-12.07	&	-12.02	\\
HeI + \fion{S}{IX}  1.252	&	-12.45	&	-12.25	&	-12.04	&	-11.74	\\
HI 1.2818	&	-13.65 &	-10.91	&	-10.91	&	-11.23	\\
PVII 1.3745	&	-14.02	&	-11.50	&	-11.50	&	-11.67	\\
\fion{Si}{X} 1.4300	&	-12.40	&	-16.20	&	-12.40	&	-11.75	\\
HI 1.6109	&	-15.36	&	-12.51	&	 -12.51&	-12.71	\\
HI 1.6407	&	-15.25 &	-12.42	&	-12.42	&	-12.60	\\
HI 1.6806	&	-15.14	&	-12.32	&	-12.32	&	-12.28	\\
PVIII 1.7393	&	-13.04	&	-12.35	&	-12.27&	-12.04	\\
HI 1.7362	&	-15.02	&	-12.21	&	-12.21	&	-12.04	\\
HI 1.8174	&	-14.88	&	-12.08	&	-12.08	&	-11.99	\\
HI 1.8751	&	-13.38	&	-10.59	&	-10.59	&	-10.81	\\
HI 1.9445	&	-14.72	&	-11.93	&	-11.93 &	-12.10	\\
\fion{Si}{VI} 1.9624	&	-15.09	&	-10.98 & -10.98 &	-11.34	\\
\fion{Al}{IX} 2.0444	&	-12.50	&	-14.48&	-12.50	&	-12.07	\\
HeI 2.0581	&	-19.50	&	-12.07	&	-12.07&	-12.07	\\
HI 2.1655	&	-14.54	&	-11.76	&	-11.75	&	-11.97	\\
HeII 2.1884	&	-14.82	&	-12.97	&	-12.96	&	-12.73	\\
\fion{Ca}{VIII} 2.3211	&	-13.34	&	-14.47 &	-13.31 &	-12.38	\\
\fion{Si}{VII} 2.4801	&	-13.16 &	-11.26	&	-11.26	&	-11.19	\\
\fion{Al}{V} 2.9045	&	-17.84	&	-11.82	&	 -11.82	&	-11.95	\\
\fion{Mg}{VIII} 3.0276	&	-11.72	&	-13.02 &	-11.70 &	-11.11	\\
\fion{Ca}{IV} 3.2061	&	-19.36	&	-12.40	&	-12.40	&	-12.31	\\
HeII 3.0908	&	-14.22	&	-12.34	&	-12.33 &	-12.07	\\
\fion{Al}{VI} 3.6593	&	-14.88	&	-12.02	&	-12.02	&	-11.88	\\
AlVIII 3.6972	&	-13.06 &	-13.72 &	-12.97 &	-12.49	\\
HI 3.7395	&	-15.00	&	-12.21 &	-12.21 &	-12.34	\\
\fion{Si}{IX} 3.9282	&	-11.98	&	-14.95 &	 -11.98 &	-11.67	\\
HI 4.0511	&	-14.08	&	-11.33	&	 -11.33 &	-11.60	\\
\enddata
\tablecomments{ \\
$^{\ddagger}$ All Cloudy input parameters are identical to the cylindrical model.\\
$^{\dagger}$ LD/HD = low/high density. }
\end{deluxetable*}

%
\begin{deluxetable*}{@{\extracolsep{0pt}}lcccc}
\tablenum{4}
\setlength{\tabcolsep}{3pt} 
\tablecaption{V1716 Sco Observed and CLOUDY Line Luminosities for the Cylindrical Model 
where LD is expanding with a higher velocity than the HD \label{tab:table-4-ldhd}}
\tablehead{
\colhead{} & \colhead{log flux\tablenotemark{$\dagger$} } & \colhead{log flux\tablenotemark{$\dagger$}} & \colhead{log total flux}  & \colhead{Dereddend} \\  
\colhead{Line} & \colhead{(LD)} & \colhead{(HD)} & \colhead{(HD + LD)} & \colhead{Observed Flux}\\
\colhead{($\mu$m)} & \colhead{(erg s$^{-1}$ cm$^{-2}$)} & \colhead{(erg s$^{-1}$ cm$^{-2}$)} & \colhead{(erg s$^{-1}$ cm$^{-2}$)} & \colhead{(erg s$^{-1}$ cm$^{-2}$)}
}
\startdata
\fion{O}{II} 0.7320/30	&	-20.13	&	-11.10	&	-11.10 &	-11.16	\\
OI 0.8446	&	$<$-20.00	&	-11.62	&	-11.62&	-11.77	\\
HI 0.9229	&	-13.95	&	-11.86	&	-11.86&	-11.69	\\
HI 0.9545	&	-13.80	&	-11.72	&	-11.71&	-11.75	\\
\fion{S}{VIII} 0.9914	&	-11.90	&	-11.47	&	-11.33  &	-11.50	\\
HI 1.0049	&	-13.62	&	-11.54	&	-11.54	&	-11.24	\\
\fion{N}{I} 1.0404	&	$<$-20.00	&	-11.50	&	-11.50&	-11.54	\\
HeI 1.0830	&	-14.99	&	-10.14	&	-10.14 &	-10.30	\\
HI 1.0938	&	-13.40	&	-11.33	&	-11.33 &	-11.27	\\
OI 1.1287	&	$<$-20.00	&	-11.75	&	-11.75	&	-11.92	\\
HeII 1.1626	&	-13.38	&	-12.09 &	-12.07	&	-12.02	\\
HeI + \fion{S}{IX}  1.252	&	-11.92	&	-12.36	&	-11.78	&	-11.74	\\
HI 1.2818	&	-13.12 &	-11.07	&	-11.07	&	-11.23	\\
PVII 1.3745	&	-13.42	&	-11.50	&	-11.50	&	-11.67	\\
\fion{Si}{X} 1.4300	&	-11.94	&	-16.21	&	-11.94	&	-11.75	\\
HI 1.6109	&	-14.80	&	-12.66	&	 -12.66&	-12.71	\\
HI 1.6407	&	-14.69 &	-12.57	&	-12.57	&	-12.60	\\
HI 1.6806	&	-14.57	&	-12.47	&	-12.47	&	-12.28	\\
PVIII 1.7393	&	-12.46	&	-12.35	&	-12.10&	-12.04	\\
HI 1.7362	&	-14.46	&	-12.36	&	-12.36	&	-12.04	\\
HI 1.8174	&	-14.32	&	-12.23	&	-12.23	&	-11.99	\\
HI 1.8751	&	-12.85	&	-10.76	&	-10.76	&	-10.81	\\
HI 1.9445	&	-14.16	&	-12.08	&	-12.08&	-12.10	\\
\fion{Si}{VI} 1.9624	&	-14.50	&	-10.98 & -10.98 &	-11.34	\\
\fion{Al}{IX} 2.0444	&	-11.96	&	-14.48&	-11.96	&	-12.07	\\
HeI 2.0581	&	-19.04	&	-12.17	&	-12.17&	-12.07	\\
HI 2.1655	&	-13.98	&	-11.91	&	-11.91	&	-11.97	\\
HeII 2.1884	&	-14.25	&	-12.97	&	-12.95	&	-12.73	\\
\fion{Ca}{VIII} 2.3211	&	-12.71	&	-14.47 &	-12.70 &	-12.38	\\
\fion{Si}{VII} 2.4801	&	-12.59 &	-11.27	&	-11.25	&	-11.19	\\
\fion{Al}{V} 2.9045	&	-17.24	&	-11.84	&	 -11.84	&	-11.95	\\
\fion{Mg}{VIII} 3.0276	&	-11.15	&	-13.02 &	-11.14 &	-11.11	\\
\fion{Ca}{IV} 3.2061	&	-18.70	&	-12.41	&	-12.41	&	-12.31	\\
HeII 3.0908	&	-13.65	&	-12.34	&	-12.32 &	-12.07	\\
\fion{Al}{VI} 3.6593	&	-14.28	&	-12.02	&	-12.02	&	-11.88	\\
AlVIII 3.6972	&	-12.49 &	-13.72 &	-12.47 &	-12.49	\\
HI 3.7395	&	-14.45	&	-12.37 &	-12.37 &	-12.34	\\
\fion{Si}{IX} 3.9282	&	-11.46	&	-14.96 &	 -11.46 &	-11.67	\\
HI 4.0511	&	-13.54	&	-11.50	&	 -11.49 &	-11.60	\\
\enddata
\tablecomments{ \\ $^{\dagger}$ LD/HD = low/high density.}
\end{deluxetable*}


\vspace{-2.2cm}
\subsubsection{Photoionization Models - Goodness of Fit} 

The $\chi^{2},$ the goodness of fit of a model to the observed spectrum, is determined by the 
following relation,

\begin{equation}
\chi^{2} = \sum_{i=1}^{n} \frac{ ( M_{i}  - O_{i} )^{2} }{\sigma_{i}^{2} } 
\end{equation}

\noindent where $n$ is the number of emission lines used in the model, $M_{i}$ and 
$O_{i}$ are the modeled and observed line flux ratios (line fluxes were normalized with respect 
to the Paschen-$\beta$  line), and $\sigma_{i}$ is the error in the observed line flux ratios. 
The reduced $\chi^{2}$ (reported in Table~\ref{tab:table-1}) is given by $\chi^{2}_{\rm{red}} = \chi^{2}/\nu,$
where $\nu$ is the number of degrees of freedom, given by the difference between the
number of observed emission lines ($n$) and the number of free parameters ($n_{p}$),
where $\nu = n - n_{p}.$ For acceptable fitting, the value of $\chi^{2} \sim \nu$ and 
$\chi^{2}_{\rm{red}}$ should be low, typically between 1 and 2. The value of $\sigma$ generally 
lies in the range of 10\% to 50\% \citep[e.g.,][]{2005ApJ...624..914V, 2010AJ....140.1347H, 2024MNRAS.527.1405H}
depending on several factors including (a)~uncertainties in the dereddening value
which is wavelength dependent (it is as large as 20\% near 0.7~\micron, 
for an error of 0.1 in the \EBminusV \, value), (b)~the actual measurement of the line 
fluxes, (c)~the process of removing the H lines from the standard A0V star spectrum before 
ratioing the nova spectrum (the H lines are numerous, consisting of lines from the Paschen, 
Brackett, Pfund and Humphreys series),  and (d)~blending of lines which is a major source of 
uncertainty in many cases. Given these factors, we consider $\sigma = 35$\% for the present
study (25\% for each line measurement and hence 35\% for each line ratio relative to
Paschen-$\beta$ 1.2818~\micron; the error being added in quadrature while ratioing).

In our analysis the total number of lines is 39, the number of free parameters = 16 and 
hence the degrees of freedom are 23. We thus get a $\chi^{2} =  59.8$ and a reduced 
$\chi^{2}_{\rm{red}} = 2.6$. A major part of the  $\chi^{2}$ value comes from 2 lines,
the \fion{O}{II} 0.7320,\,30~\micron{} line ($\chi^{2} = 7.7$) and from the \fion{Si}{VI} 1.96~\micron{}
line ($\chi^{2}= 12.8$). If these two lines are omitted, $\chi^{2}_{\rm{red}} \simeq 1.87.$ 
This is acceptable especially since the assumption of a cylindrical geometry 
is a significant simplification \citep[e.g.,][]{2022ApJ...925..187P, 2024MNRAS.527.1405H}.

\begin{figure}[htb!]
\figurenum{5}
\begin{center}
\includegraphics[trim=0.4cm 0.5cm 0.25cm 0.15cm, clip, width=0.43\textwidth]{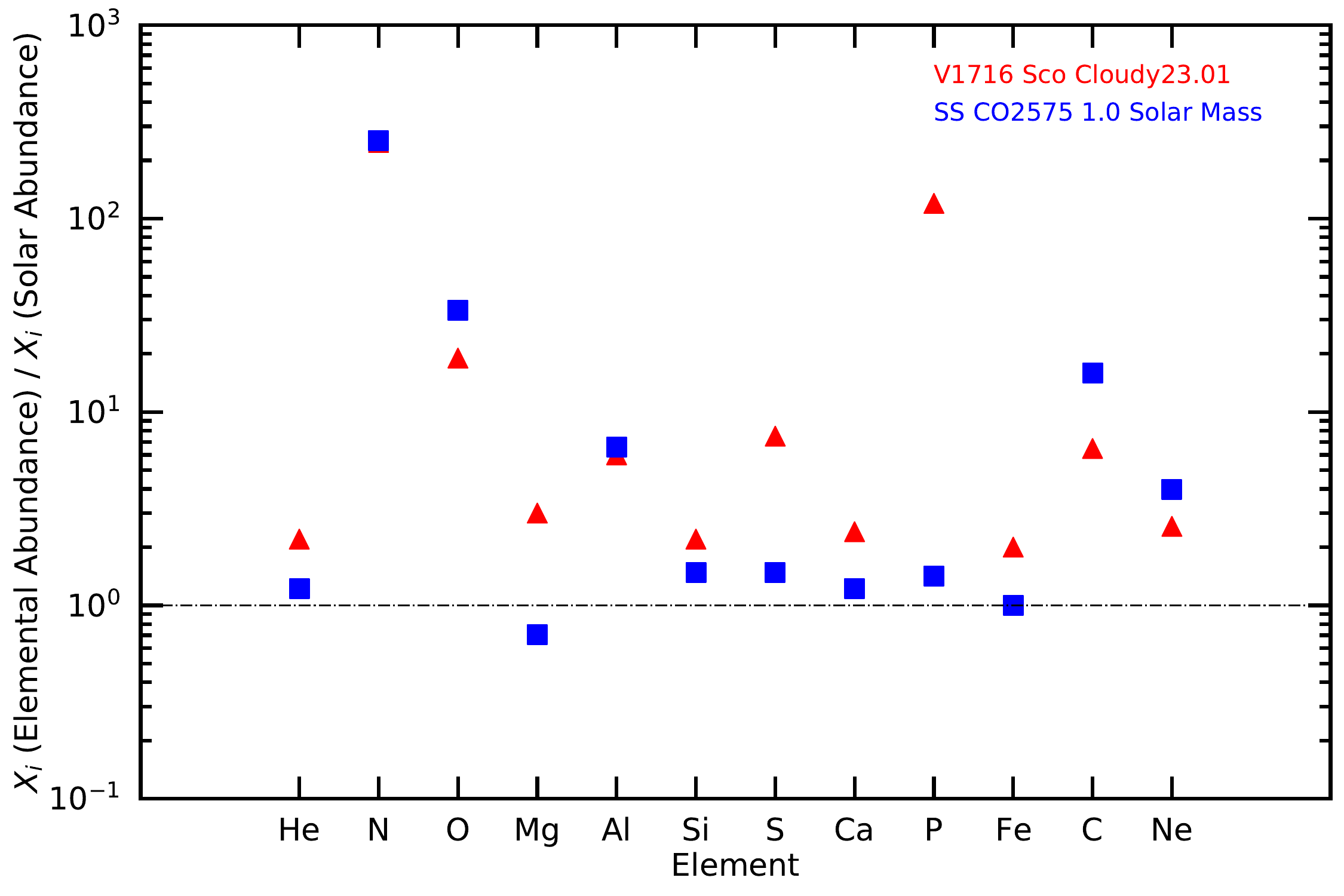}
\caption{The observed abundances (red triangles) derived from the Cloudy photoionization model (assuming a cylindrical
geometry) versus model abundance values obtained from tabulated mass fractions \citep[blue squares,][] {2020ApJ...895...70S}
for a 1.0~\Msun{} CO nova with 25-75\% mixing (see Section~\ref{sec:sec-abund} for further details). Solar abundance values are taken 
from \cite{2010ApSS.328..179G}. The dashed line across the plot represents solar abundances. The blue square for Fe is set to
the solar abundance value as the temperatures reached in the model TNRs are insufficient to produce Fe. The TNR event itself
removes the original abundance heritage of the material accreted onto the WD surface by explosive nucleosynthesis reactions 
\citep{2020ApJ...895...70S, 2024ApJ...962..191S}. 
}
\label{fig:fig-CvsSS-Abundance} 
\end{center}
\end{figure}

\vspace{0.25cm}
\subsection{Abundances Estimates and WD Ejecta Mixing}
\label{sec:sec-abund}

A comparison of the observed line intensities with the co-added CLOUDY line intensities 
(the fluxes of high and low density components are co-added) is presented in Figure~\ref{fig:fig-threeA}.
Considering that a variety of lines are seen (recombination lines, lines from 
neutral species, lines from highly ionized atoms, Ly$\beta$ fluoresced lines), a reasonably 
good reproduction of line strengths is seen in Figure~\ref{fig:fig-threeA}. 
H is depleted from nuclear burning in the TNR. We find (Figure~\ref{fig:fig-CvsSS-Abundance}) that the 
observationally derived abundances of He, Ne, C, O, Fe, Al, 
Si, S and Ca are mildly to moderately above solar. The abundances (by mass) with respect
to the Sun \citep{2010ApSS.328..179G} are  He = 2.20, C = 6.47, O = 18.99, Ne = 2.56, Mg = 3.00, Al = 6.00, Si = 2.20, 
S = 7.50, Ca = 2.42, Fe = 2.00, N = 248.00 and P = 120.0.  Notably, phosphorus and nitrogen are highly overabundant.
However, this is consistent with the predictions of \citet{2020ApJ...895...70S, 2024ApJ...962..191S} wherein 
P is predicted to be 100 or more times overabundant (compared to solar) for massive ONe or CO WDs. 
For example, for a 1.25~\Msun{} CO WD with 25-75\% mixing the value of P is $\sim 100.$
The P overabundance may be suggesting that V1716~Sco harbors a massive WD ($\gtsimeq 1.0$~\Msun)
which is consistent with its short $t_{2}$, (see Section~\ref{sec:sec-properties}). 

We have compared the derived abundances with those expected in the nucleosynthesis models of 
\citet{2020ApJ...895...70S} as illustrated in Figure~~\ref{fig:fig-CvsSS-Abundance}.
This figure compares the observed CLOUDY deduced yields in V1716~Sco 
with one of the \citet{2020ApJ...895...70S} models of an 1.0~\Msun{} WD with a CO core 
with 25-75\% mixing. A mixing ratio of 25-75\% means that the material that 
undergoes a TNR has a composition of 25\% of the outer layers of the underlying 
WD mixed with 75\% of the accreted envelope during the to thermonuclear 
runaway. The agreement is reasonable for many of the elements.

\begin{figure}[ht!]
\figurenum{6}
\begin{center}
\includegraphics[trim=0.05cm 0.05cm 0.05cm 0.05cm, clip, width=0.43\textwidth]{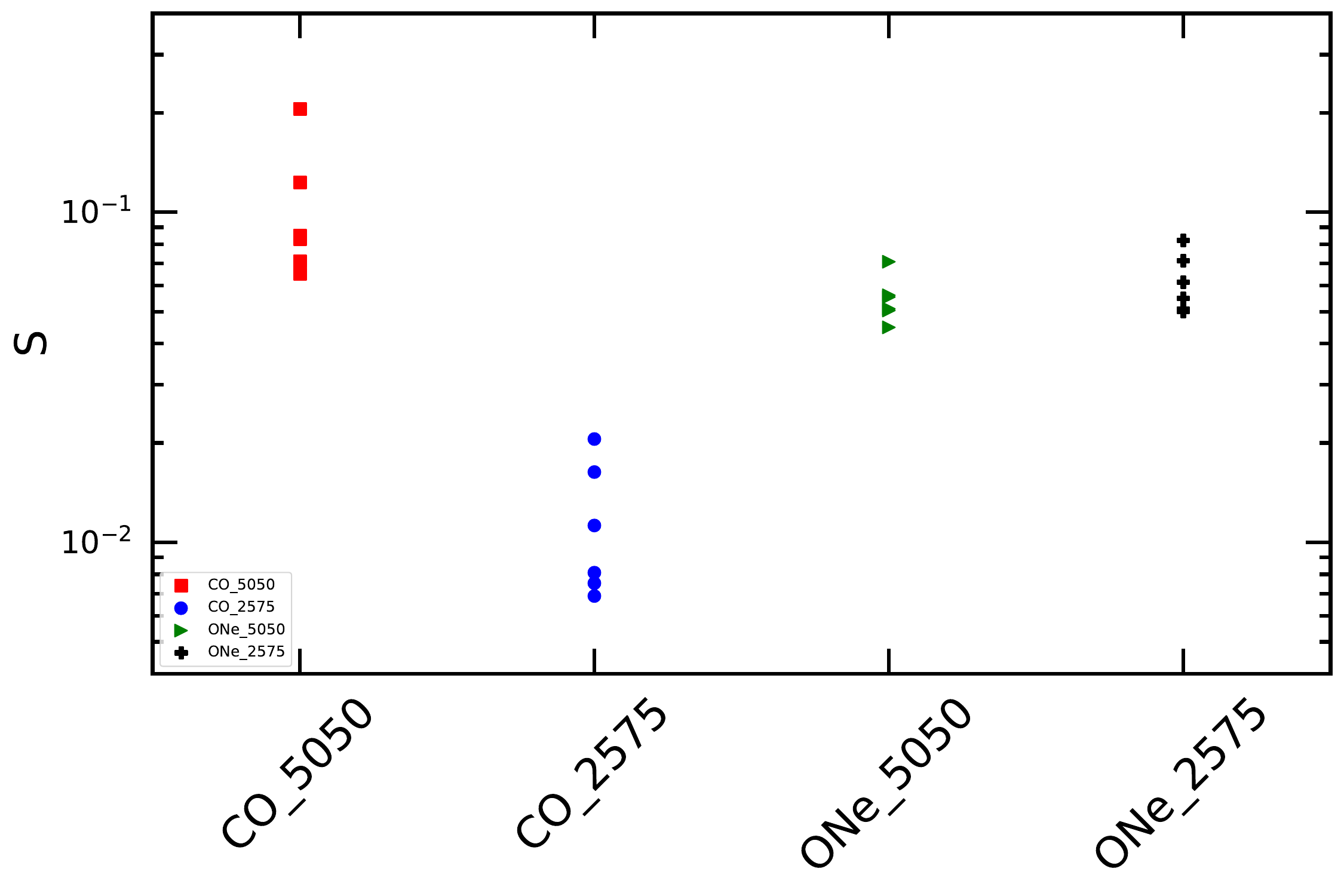}
\caption{Comparison of TNR mixing values to derived abundances in V1716~Sco. 
Each tick label on the X-axis (e.g., CO\_\,2575) gives the model values taken from
\citet{2020ApJ...895...70S, 2024ApJ...962..191S}, the type of the WD (CO or 
ONe) and the mixing fraction (e.g., 25-75\%). The multiple points along the Y axis 
for each tick label correspond to the $S$ values for the different 6 WD masses 
(0.6, 0.8, 1.0, 1.15, 1.25, and 1.35~\Msun)
considered in the latter references. $S$  is essentially a meter to broadly measure 
how well the observed abundances in V1716~Sco match  with nucleosynthesis
models (see Section~\ref{sec:sec-abund}).  A small value of $S$ implies a 
good agreement. } 
\label{fig:fig-fiveA} 
\end{center}
\end{figure}

It is not known whether V1716~Sco contains a WD of the CO or ONe type. The mass of the 
WD is also not known. However the Fe II classification of the nova \citep{2023ATel16003....1W} 
strongly suggests a CO type.  Figure~\ref{fig:fig-fiveA} shows a comparison of different models of 
\citet{2020ApJ...895...70S, 2024ApJ...962..191S} for both CO and ONe novae 
with different WD masses and mixing fractions. For each of the CO and ONe classes, 
there are six Starrfield models shown with WD masses of 0.6, 0.8, 1.0, 1.15, 1.25 and 1.35~\Msun{} 
respectively. For each WD mass, two mixing fractions are considered that of 25-75\% and 50-50\%. 
Thus there are a total of twenty-four \citet{2020ApJ...895...70S, 2024ApJ...962..191S} models.

\citet{2013ApJ...777..130K} investigated whether or not observed ONe nova abundances can be used 
to constrain the degree of mixing that occurs between the outer layers of the underlying WD
and the accreted envelope prior to TNR. Any abundance used for this purpose, was 
referred to as a mixing meter. Comparison of mixing meters with observations allowed for an
estimate of the mixing fractions in individual novae. They found a fraction of 25\% or smaller 
for the mixing between WD matter and the accreted envelope in almost all cases (ONe models). 
Therefore, \citet{2013ApJ...777..130K} concluded that the observations support a mixing 
fraction that is much smaller than 50\%, which has usually been used in the literature.

In Figure~\ref{fig:fig-fiveA}, the meter used is the quantity $S$ defined as

\begin{equation}
S = \frac{1}{N} \sum_{i=1}^{n}  \Bigl(\frac{\rm{x}_{obs}(i)}{\rm{x}_{\odot}} - \frac{\rm{x}_{model}(i)}{\rm{x}_{\odot}}\Bigr)^{2}
\end{equation}

\noindent where the summation is over $N$ (= 11), the number of elements with derived CLOUDY abundances
whose mass fractions are compared to those derived for the ejecta for various WD mass in the
TNR models of  \citet{2020ApJ...895...70S, 2024ApJ...962..191S}. On the X-axis, each 
tick label (e.g., CO\_\,2575)  gives the the type of the WD (CO or ONe) and the 
mixing fraction (e.g., 25-75\%). 

Comparing the CO\_\,5050 and CO\_\,2575 values in Figure~\ref{fig:fig-fiveA}, it is clear that the 
latter set has much smaller scatter and smaller $S$ values strongly suggesting that a 
25-75\% mixing fraction is certainly favored over a 50-50\% mixing for V1716~Sco. 
Further, inter-comparison between the SS ONe and CO models, shows that the observed 
abundances in V1716~Sco match the CO novae better. A similar conclusion is reached 
comparing to models by \citet{2020A&A...634A...5J}. 
This suggests that V1716~Sco is a CO nova, an inference that is consistent with the CO classification 
proposed in Section~\ref{sec:sec-properties},  based on the Fe~II spectral classification.
Absence of neon lines in the spectra at day 133 also suggests that the binary contains a CO WD, although
late-time observations when the source enters a nebular phase combined with detailed abundance
modeling may be necessary to confirm the latter assertion 
\citep[e.g., see the cautionary note in][]{2007AJ....134..516S} as \fion{Ne}{V} 3550/3426~\AA{} was
detected by Swift near the onset of the SSS phase \citep{2023ATel16069....1P}. 

As a check for the sensitivity of the results of Figure~\ref{fig:fig-fiveA} to errors in the observed flux, 
we ran an additional two sets of computations with the abundances of all the elements 
under consideration uniformly increased (or decreased) by $\pm$ 20\% with respect to H. 
No significant changes are seen to the results presented in Figure~\ref{fig:fig-fiveA}.

\subsection{Ejecta Mass}
\label{sec:sec-ejmass}

The mass  (M$_{\rm{shell}}$) in a cylindrical volume of height $h$ can be calculated 
in units of \Msun{} from the parameters presented in Table~\ref{tab:table-1} by the expression:

\begin{equation}
\begin{aligned}
\rm{M}_{\rm{shell}} = \pi \, m_{p} \, (\rm{r}_{\rm{out}}^2 - \rm{r}_{\rm{in}}^2)\, \rm{h}\,
(f_{\rm{HD}} n_{\rm{HD}} + f_{\rm{LD}} n_{\rm{LD}} ) \sum (1 + \rm{A}_{i}) \\
\end{aligned}
\end{equation}

\vspace{0.1cm} 
\noindent where $f_{\rm{HD}}$, $n_{\rm{HD}}$, $f_{\rm{LD}}$, $n_{\rm{LD}}$ 
are the filling factors and H densities for the high (``clumps'') and low density 
(``diffuse'') components  (abbreviated as HD and LD respectively) given in 
Table~\ref{tab:table-1}, $m_{p}$  is the proton mass, and the summation is over 
the fractional mass of the elements in the shell with respect to hydrogen (using 
the fraction-by-numbers data in Table~\ref{tab:table-1}; other elements not listed in Table~\ref{tab:table-1} 
are assumed to have solar values). We obtain a mass of $\simeq 4.19\times 10^{-4}$~\Msun, (adopting
average values of $n_{\rm{HD}}$ and $n_{\rm{LD}}$ assuming $n(r) = n(r_{\rm{in}}) \times r^{-3}$
where $\sim 88$\% is in the HD component and $\sim 12$\% is in the LD component.
This ejected mass estimate is greater than those predicted from TNR models. However,
this is a well known tension that has plagued studies of the nova phenomena for
decades \citep[e.g.,][]{1999PhR...311..371S, 2011CaJPh..89..333W}. The underlying 
cause for this discrepancy is not understood.

\begin{figure}[ht!]
\figurenum{7}
\begin{center}
\includegraphics[trim=0.05cm 0.55cm 0.55cm 1.25cm, clip, width=0.42\textwidth]{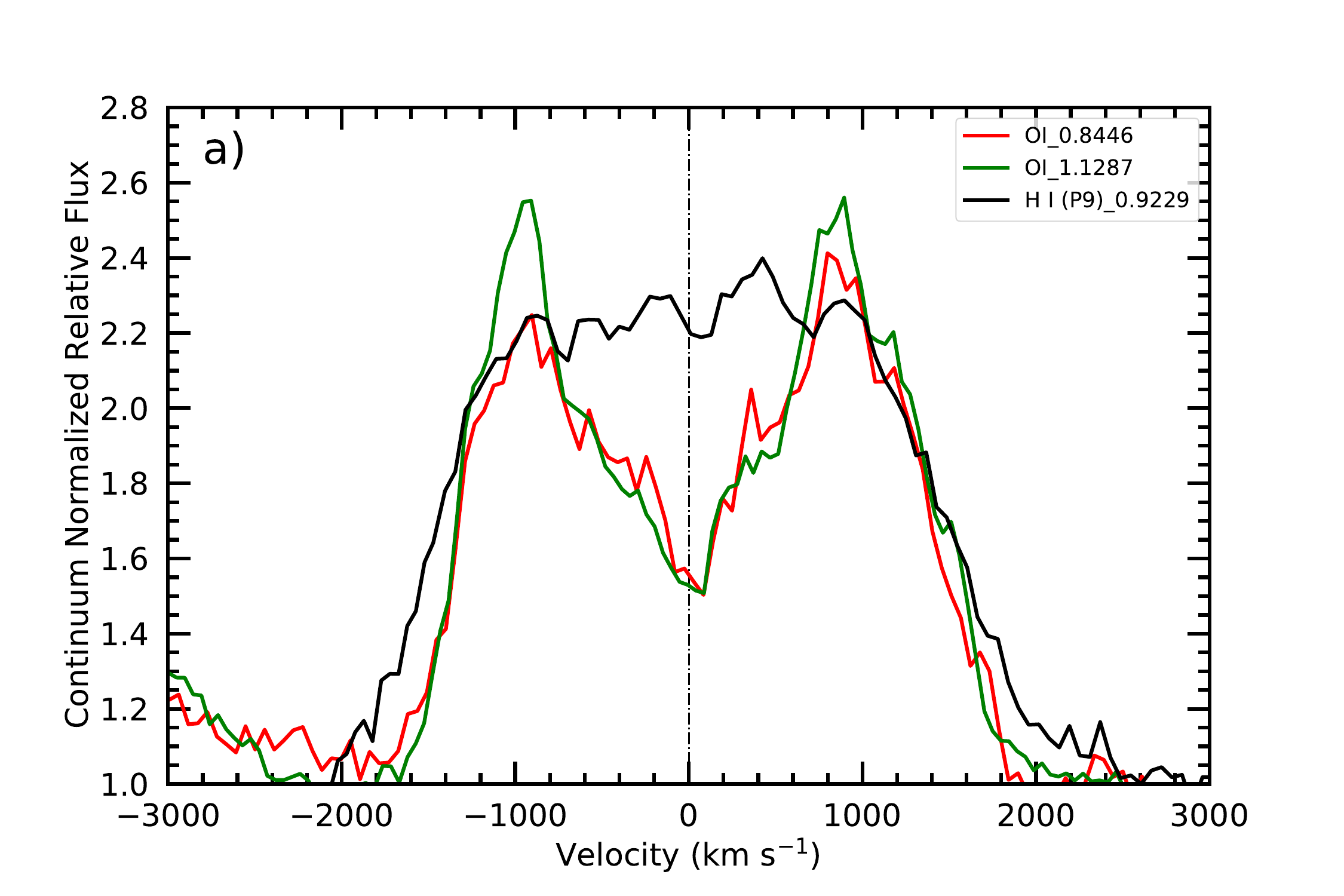}
\includegraphics[trim=0.05cm 0.55cm 0.55cm 1.25cm, clip, width=0.42\textwidth]{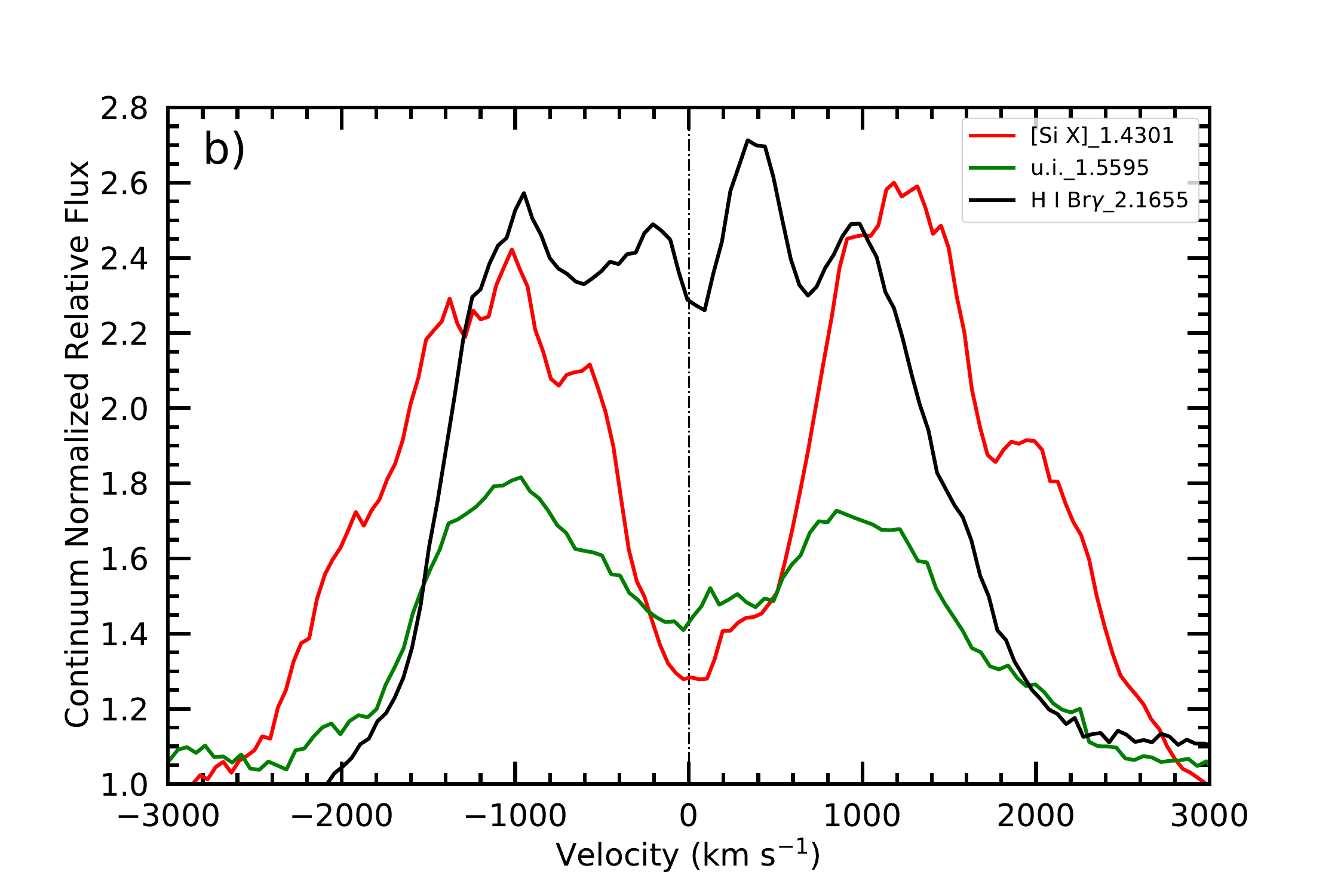}
\includegraphics[trim=0.05cm 0.55cm 0.55cm 1.25cm, clip, width=0.42\textwidth]{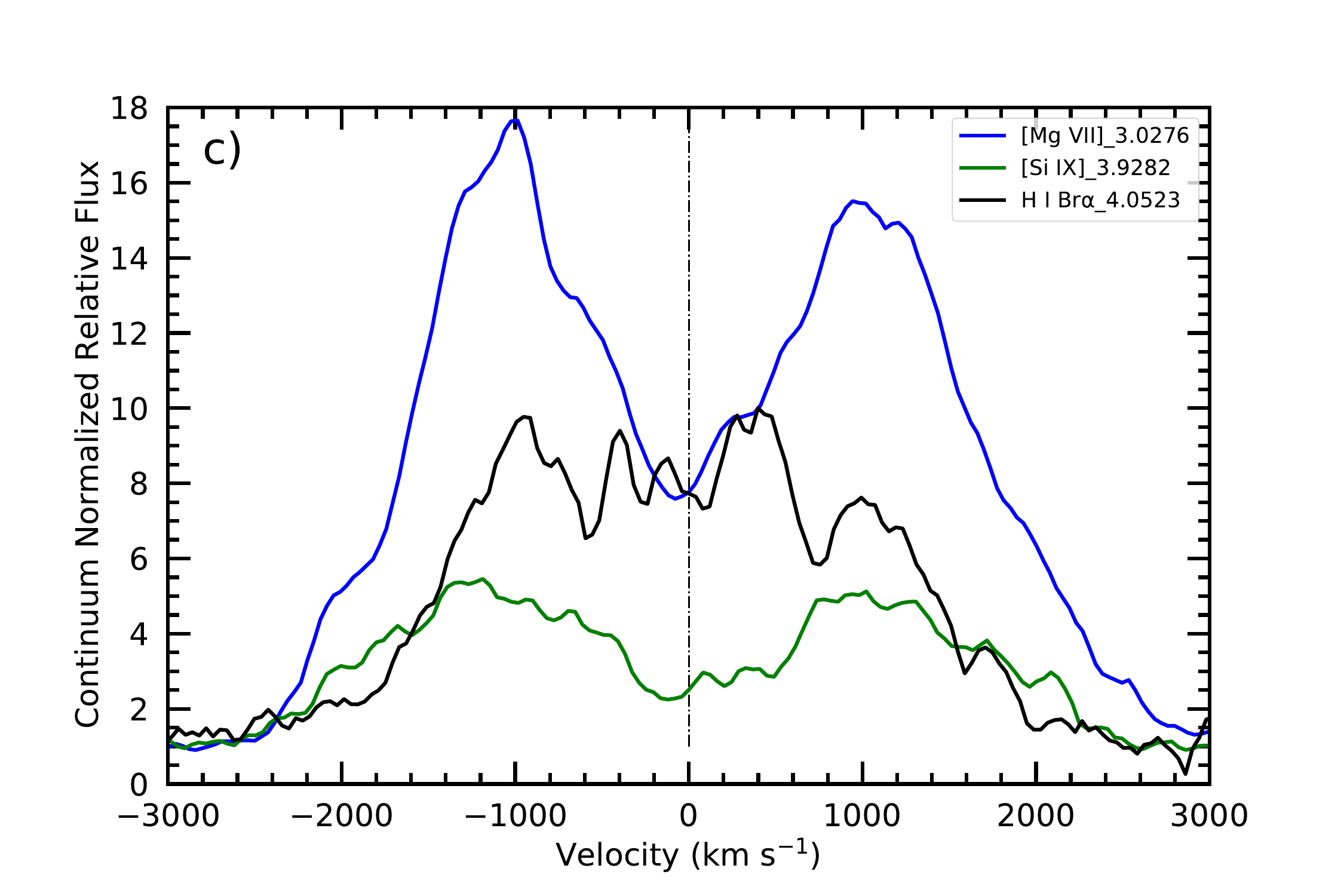}
\includegraphics[trim=0.05cm 0.55cm 0.55cm 1.25cm, clip, width=0.42\textwidth]{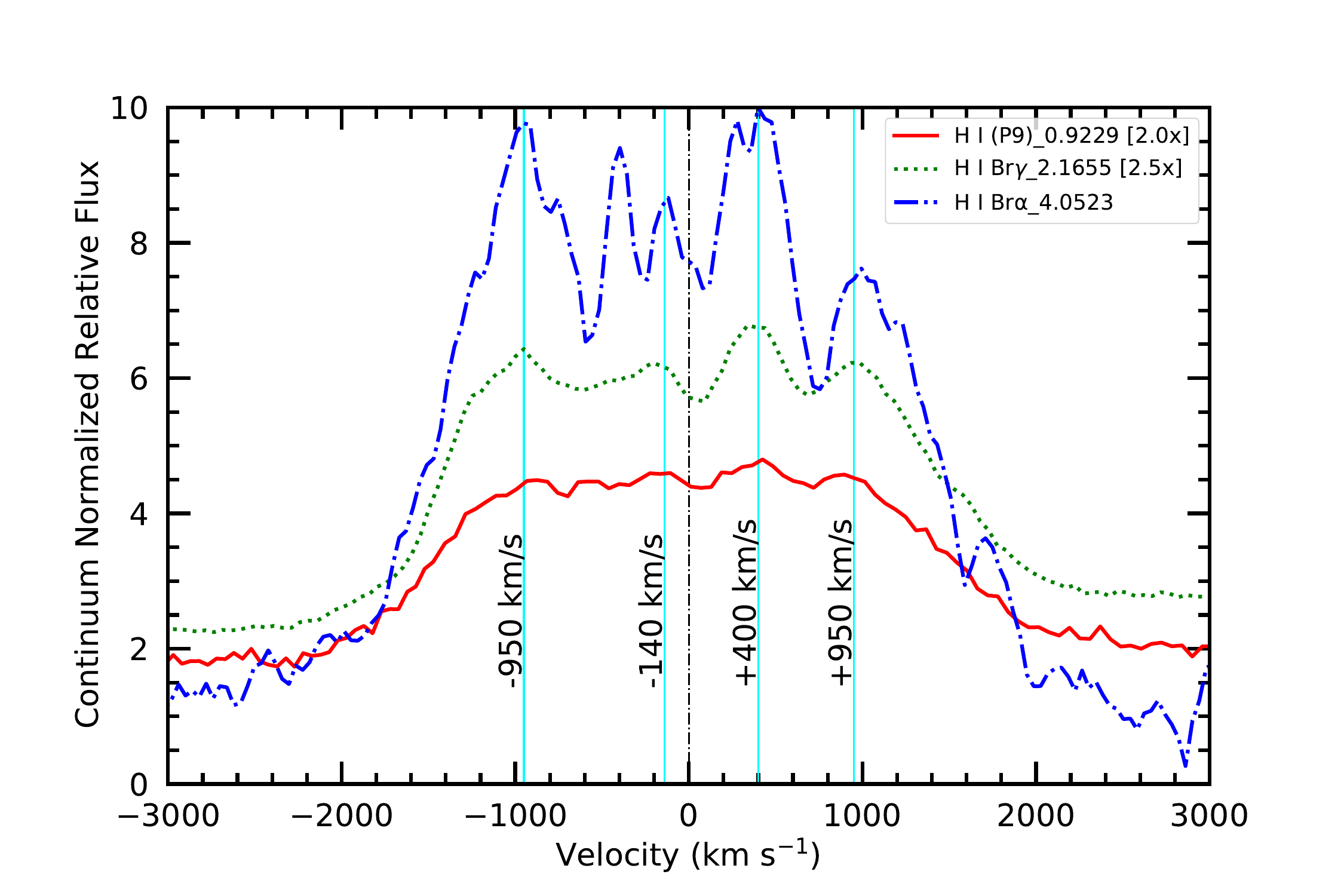}
\caption{Emission line velocity profiles of V1716~Sco on day +132.8.  
Each emission line is normalized to the average adjacent continuum and
scaled (as indicated in the inset). The vertical black dashed line in each panel is 0 km~s$^{-1}$,
and the `u.i.' label indicates unidentified line.   
a)~Lines of neutral oxygen O~I 0.8446 and 1.1287~\micron{} versus H~I (P9) 0.9229~\micron.  
b)~Coronal lines of \fion{Si}{IX} versus H~I Br$\gamma$ in the SpeX SXD mode. 
c)~Coronal lines \fion{Mg}{VII} and \fion{Si}{IX} in the thermal 
IR ($\lambda \gtsimeq$ 3.0~\micron) versus H~I Br$\alpha$ in the SpeX LXD\_\,short 
mode. The FWHM of the \fion{Mg}{VIII} profile is $\simeq 3400$~km~s$^{-1}$. d) Common
velocity components present in the H~I recombination line profiles (vertical black arrows). 
}  
\label{fig:fig-7velocities} 
\end{center}
\end{figure}

\vspace{0.5cm}
\section{Discussion}
\subsection{Interpretation of Line Profiles and CLOUDY}
\label{sec:sec-profiles}

There is a clear difference between the shape of the line profiles of the H and He lines, 
both of which arise from the dense component, and the coronal lines (with ionization
potentials of the lower ionization state $\gtsimeq 225$~ev),  which mostly 
arise from the low density component (Table~\ref{tab:table-2-cylindar}). This is illustrated
in Figure~\ref{fig:fig-7velocities}.

H and He lines are broad with a castellated peak 
emission profile, whereas the infrared coronal lines exhibit distinct double-peaked 
structures with a deep dip between the peaks (a notched saddle profile) centered
near $\simeq 0$~km~s$^{-1}$. The lines showing this structure are listed in 
Table~\ref{tab:table-5-peaks} along with the peak-to-peak separation.
The highest velocity, common H~I recombination line components (Figure~\ref{fig:fig-7velocities}d)
at $\pm 950$~km~s$^{-1}$ are coincident with the double peak profiles seen in neutral 
oxygen, while the coronal line double peaks are at higher velocities $\pm 1100$~km~s$^{-1}$ .
Wings of the coronal lines exhibit even higher velocity substructures in their profiles. 
The FWHM of the \fion{Mg}{VIII} profile (Figure~\ref{fig:fig-7velocities}c) is $\simeq 3400$~km~s$^{-1}$. 

The velocity structures and FWHM suggest that the lines arise from different regions 
within the ejecta. This is expected because the forbidden coronal lines will be collisionally 
de-excited in a region that is denser than the critical density of the 
line \citep{2021ApJ...922L..10W, 2022MNRAS.510.4265K}.
What is puzzling is that the O~I 0.8446 and 1.1287~\micron{} line profiles \citep[excitation 
potential 12.03 ev,][]{1950ApJ...111....1K, 1995ApJS...96..325B} also show a similar profile 
(albeit somewhat narrower in peak-to-peak separation) as the coronal lines. Modeling indicates
(Table~\ref{tab:table-2-cylindar}) oxygen lines also arise within denser regions of the ejecta.

A significant part of the strength of both the 0.8446 and 1.1287~\micron{} 
lines is from Lyman-$\beta$ fluorescence. So it is consistent that the sites of 
O~I and H emission are co-spatial (Figure~\ref{fig:fig-7velocities}a) ensuring an adequate input 
of Ly-$\beta$ photons from the recombination cascade process in hydrogen. If the H emitting 
region is optically thick, Ly-$\beta$ photons may even get trapped. Thus the necessary 
sources of Ly-$\beta$ photons are available for fluorescing the O~I lines. However, the O~I lines 
show a considerably smaller expansion velocity (FWHM $\simeq 1350$~km~s$^{-1}$) than the 
other coronal lines (FWHM $\gtsimeq 3000$~km~s$^{-1}$) strongly suggesting that they originate from 
different regions despite sharing the same double-peaked profile shape.

Likely the V1716~Sco ejecta nebula is not spherically symmetric, but bipolar in morphology.
Bipolar morphologies are  directly detected in several novae,  such as 
V1280~Sco \citep{2012A&A...545A..63C, 2022ApJ...925..187P},
RS Oph \citep{2007ApJ...665L..63B},  V959 Mon \citep[][and references therein]{2017MNRAS.469.3976H}
and V445 Pup \citep{2009ApJ...706..738W, 2021MNRAS.501.1394N}, by using high spatial resolution 
imaging techniques and interferometry;  hence this geometry is common (and why 
Cloudy models in Section~\ref{sec:sec-cloudy} explore cylindrical geometries). 

In V1716~Sco, denser regions in the asymmetrical ejecta could be the site of the 
O~I 0.8446/1.1287~\micron{} lines and the O~I double-peaked profiles could arise 
from the receding and approaching parts of these lobes. The strong shock created by 
the ejecta colliding with dense globules, or denser toroidal material within 
the ejecta could likely have given rise to the $\gamma$-rays seen during the 
outburst.  The geometric scenario for the $\gamma$-ray 
generation proposed here would then be similar to that proposed for the 
$\gamma$-ray nova V959~Mon \citep{2014Natur.514..339C}. In comparison to the O~I 
lines, the coronal lines with a higher velocity (Table~\ref{tab:table-5-peaks}), could 
originate from the bipolar lobes which are expected to be more diffuse and also to have 
a higher velocity compared to the material in a constricting denser torus. Some of 
the coronal lines with high critical densities in the range of $10^{8}$ to $10^{9}$~cm$^{-3}$, 
like the \fion{Si}{VI} 1.96 and \fion{Si}{VII} 2.48~\micron{} lines 
for gas temperatures in excess of $10^{5}$~K, \citep{2023MNRAS.522.4841E}
could come from both the torus and bipolar lobes. Hence their line profiles are expected 
to be as evident in Figure~\ref{fig:fig-7velocities}.  

\begin{figure}[hbt!]
\figurenum{8}
\begin{center}
\includegraphics[trim=0.01cm 0.01cm 0.01cm 0.01cm, clip, width=0.42\textwidth]{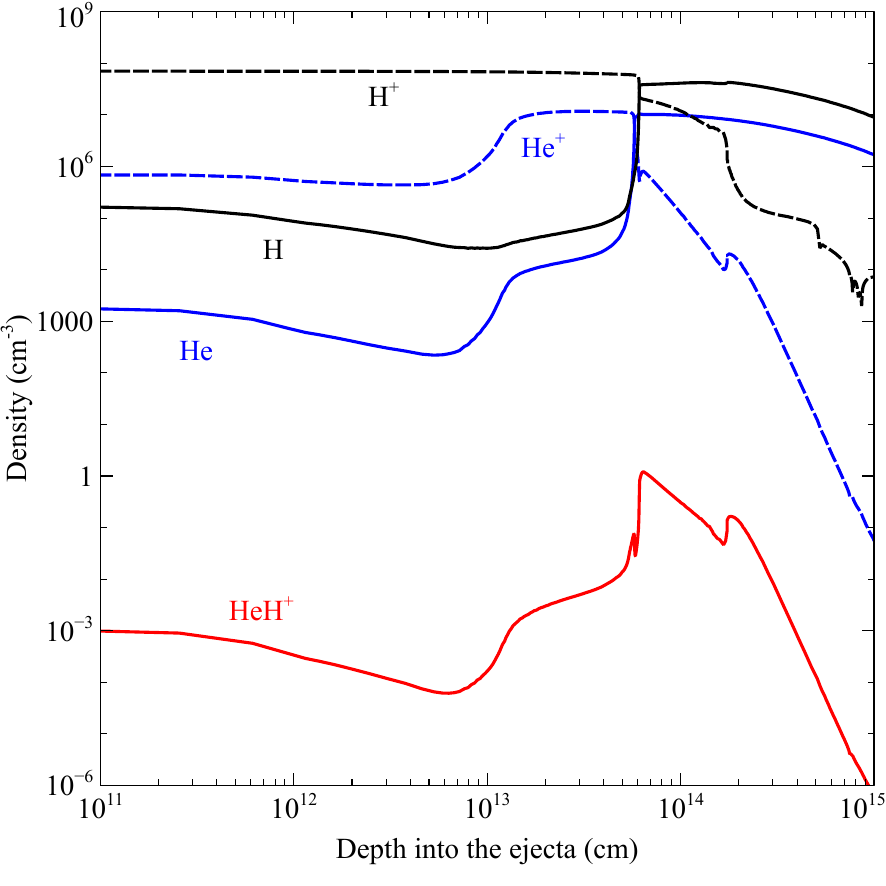}
\caption{CLOUDY derived number densities (cm$^{-3}$) of HeH$^{+}$ (red) and that of of He (blue) 
and H$^{+}$ (black) as a function of depth into the nova ejecta (cm) in V1716~Sco (+133~d). The ionization
front is near $\simeq 6 \times 10^{13}$ cm.}
\label{fig:fig-heh} 
\end{center}
\end{figure}

The \fion{N}{I} 0.5755~\micron{} and 1.04~\micron{} lines, together with the \fion{O}{I} 0.6300, 0.6364~\micron{} 
doublet, are among the first forbidden lines to appear in the spectra of novae (particularly the Fe~II 
novae) and they remain present even after high ionization lines appear in the spectra \citep{1994ApJS...90..297W,  2024MNRAS.527.9303A}.
The \fion{N}{I} 1.04~\micron{}  line is seen here and inspection of optical spectra from the ARAS 
database\footnote{\url{http://www.astrosurf.com/aras/Aras_DataBase/Novae.htm}} shows that 
\fion{O}{I} 0.6300, 0.6364~\micron{}  lines were present before and during the current observations on day 133. 
Based on several characteristics of the \fion{O}{I} emission, \citet{1994ApJS...90..297W} 
 has convincingly argued that dense globules must exist in the ejecta and that the \fion{O}{I} 
emission comes from within these globules (these globules are likely sites for dust formation too). 
Just like the \fion{O}{I}  lines, it would appear reasonable to expect that neutral \fion{N}{I} 1.04~\micron{} 
emission also originates from material in the dense globules, where nitrogen atoms 
can remain shielded and neutral. This is supported by the data in Table~\ref{tab:table-2-cylindar} which show 
that \fion{N}{I} emission arises from the dense component and is absent in the diffuse component.
 
However, the line profiles (Figure~\ref{fig:fig-7velocities}) alternatively could be interpreted as being 
consistent with a single ejecta component (non-spherical, axisymmetric, and inhomogeneous) 
with just a density gradient imposed by the expansion velocity law. In this picture the higher velocity 
regions would inevitably be the ones with the lowest density, explaining the broader width and larger 
peak separation of the IR coronal lines. Conversely in the lower (inner, for an explosive expansion law) velocity 
regions, recombination occurs, and Ly-$\beta$  can be sufficiently optically thick to pump the observed 
O~I transition.
 
\subsection{Observed coronal lines not replicated by CLOUDY}
\label{sec:sec-lines-nocloudy}

There are two emission lines at $\sim$ 1.55 and $\sim$ 2.09~\micron{} which are often seen 
during the coronal stage in many novae, for example V1974~Cyg \citep{1996ApJ...467..860W},
RS~Oph \citep{2009MNRAS.399..357B}, V1674~Her \citep{2021ApJ...922L..10W} and
V6558~Sgr \citep{2018ApJ...858...78G}, whose identification has remained 
uncertain. These lines are thought to be coronal in nature because they arise during the 
coronal phase and their broader profile shapes (e.g., FWHM and 
peak-to-peak separations) often replicate the coronal line profiles
shapes \citep[see][for a discussion on the 2.09~\micron{} profile shape]{2018ApJ...858...78G}
rather than the H or He line profiles shapes and substructures (see Figure~\ref{fig:fig-7velocities}
for example).  

The line at $\sim$1.55~\micron{} (labeled `u.i.', see Figure~\ref{fig:fig-one}) has been thought 
to be either \fion{Si}{IX} 1.55995~\micron{} or \fion{Cr}{XI}~1.5518~\micron{} 
\citep{1996ApJ...467..860W}. However, our CLOUDY calculations underproduce the 
\fion{Si}{IX} 1.55995~\micron{} line by a factor of 23,000, while the observed line center
is at 1.5534~\micron{}, displaced by 0.0065~\micron{} from the expected position. It is
unlikely that this line is \fion{Si}{IX}. Similarly the observed fluxes of the \fion{Cr}{XI}~1.5518
and \fion{Mn}{XIV} 2.09~\micron{} lines in V1716~Sco cannot be matched by CLOUDY
even if both these elements are 1000 times overabundant compared to their solar values. 
It is thus likely that the \fion{Cr}{XI} and  \fion{Mn}{XIV} assignments to these lines are incorrect.
In our CLOUDY models, the strongest lines between 1.5528 - 1.5601~\micron{} are
C~IV in the low density component, and blends of He~I + H~I + C~IV in the high density
component.

\subsection{Do CNe Produce Helium Hydride (HeH$^{+}$) ?}
\label{sec:sec-heh}

The spectrum of V1716~Sco consists mainly of atomic and ionic lines. However our final 
CLOUDY model predicts an observable amount of the helium hydride ion, HeH$^+$,
in component B (denser component) at a significant column density of 10$^{12.959}$ cm$^{-2}$.  
The dominant formation channel for  HeH$^+$ is He$^+$ + H $\rightarrow$ HeH$^+$ + $\it{h \nu}$. 
Uncertainties in chemical reaction rate coefficients influence the predicted abundances/column 
densities of the species involved. We adopted the rate coefficient for this reaction from 
\citet{1990ApJ...365..239Z}.  There can be 30\% uncertainty in the predicted column density 
due to the uncertainties in the chemical rate coefficients \citep{1990ApJ...365..239Z}. 

After the Big Bang, in the early Universe's metal-free and low-density environment, the 
first molecule to form was HeH$^{+}$, following radiative association of He atoms with 
protons. HeH$^{+}$ has recently been discovered toward the planetary nebula NCG 7027 
via the rotational transition at 149.1~\micron{} \citep{2019Natur.568..357G}. 
HeH$^{+}$ in NGC~7027 has a column density that is similar to that found here for V1716~Sco.
The  conditions in planetary nebulae are known to be suitable for the production of 
potentially detectable HeH$^{+}$ column densities: the hard radiation field from the 
central hot WD creates overlapping Str\"{o}mgren spheres 
\citep{2019Natur.568..357G} where HeH$^{+}$ is predicted to form. 

Figure~\ref{fig:fig-heh} shows that a similar situation exists in V1716~Sco 
(and may possibly exist in other novae) with a similar overlapping of He$^{+}$ and 
H zones (the CLOUDY model is a time-independent model, hence these are 
overlapping Str\"{o}mgren spheres) having substantial number densities of HeH$^{+}.$ In 
the near-IR, HeH$^{+}$ has a few ro-vibrational features, e.g., the $\nu = 1 - 0$ P(1) 3.51629~\micron{} line, 
the P(2) 3.60776~\micron{} line, and the $\nu = 1 - 0$ R(0) ro-vibrational line at 3.364~\micron. 
The first two of these lines have been detected in NGC 7027 \citep{2020ApJ...894...37N}.
The line fluxes of these three lines, based on the CLOUDY number densities, need to be calculated 
theoretically (which is outside the scope of this work) to know whether they should be detectable. 
However, close examination of our spectrum of V1716~Sco reveals no significant emission at 
the wavelength positions of these lines, though the atmospheric transmission in the
spectral region covered by these lines is poor. 

The study of other CNe  with the high sensitivity achievable with the JWST 
NIRspec may unambiguously detect the presence of this important hydride, and even 
its evolution in real time or provide more definitive observational evidence that rebuts 
the conjecture.

%
\begin{deluxetable}{@{\extracolsep{0pt}}lclc}
\tablenum{5}
\setlength{\tabcolsep}{0pt} 
\tablecaption{Lines with double peaked profiles\label{tab:table-5-peaks}}
\tablehead{
\colhead{Line} & \colhead{Peak-to-Peak} & \colhead{Line} &  \colhead{Peak-to-Peak} \\  
& \colhead{separation} & & \colhead{separation} \\
\colhead{($\mu$m)} & \colhead{(km s$^{-1}$)} & \colhead{($\mu$m)} & \colhead{(km s$^{-1}$)}
}
\startdata
O I 0.8446	&	1705	&	\fion{Si}{X} 1.4300	&	2350	\\
\fion{S}{VIII} 0.9914	&	1970	&	`u.i.' 1.5599	&	2120	\\
O I 1.1287	&	1807	& \fion{Mg}{VIII} 3.0276 	&	2070	\\
\fion{S}{IX} 1.2523 	&	2275	&	\fion{Si}{IX} 3.9282 &	2288	\\
\enddata
\end{deluxetable}

\section{Conclusion}
\label{sec:sec-conclude}

A moderate resolution near-IR spectrum of V1716~Sco during the coronal line phase of evolution obtained
132.8~days post-outburst was modeled by using the photoionization code CLOUDY. Abundances were 
estimated for H, He, N, O, Si, Al, Mg, S, Ca and P. \, Except for H, the analyzed elements are 
over-abundant compared to solar values. The abundances (by mass) with respect to the 
Sun are He = 2.20, C = 6.47, O = 18.99, Ne = 2.56, Mg = 3.00, Al = 6.00, Si = 2.20, 
S = 7.50, Ca = 2.42, Fe = 2.00, N = 248.00 and P = 120.0.  

It was necessary to consider the ejecta to be composed of two components. One, a dense 
component from which the bulk of the H, He, OI and N emission arises and second, a 
less dense component from which most of the coronal lines arise. Some of the coronal 
lines are found to arise from both components. A cylindrical (non-spherical) geometry
for the ejecta in photoionization modeling best reproduces the observed IR line fluxes.
The mass of the ejecta, including neutral and ionized gas, is $\simeq 4.19\times 10^{-4}$~\Msun.

The Ly-$\beta$ fluoresced O~I 0.8446 and 1.1287~\micron{} lines exhibit a prominent
double-peaked structure; a profile shape that the O~I lines share with several coronal lines.

Finally, the derived abundance yields were compared to various simulations of the TNR event
to assess and constrain the level of potential mixing \citep{2020ApJ...895...70S, 2024ApJ...962..191S}.
Our analysis suggests that in the case of V1716~Sco (which has a CO WD), a fraction of 25\% 
rather than 50\% is  favored for the mixing between WD matter and the accreted envelope 
before the outburst. This is similar to the 25\% mixing fraction that is favored in ONe 
novae \citep{2013ApJ...777..130K} and elsewhere.

 \vspace{0.1cm}

\noindent\textbf{Acknowledgements}\, 
\vspace{0.1cm}

The authors wish to acknowledge the referee for their insight and constructive critiques that
improved the manuscript. The authors also are deeply indebted to Dr. D.~P.~K. Banerjee
for his in depth discussion that paved the way for the analysis and interpretation of these data.
We also acknowledge with thanks the variable star observations 
from the \textit{AAVSO International Database} contributed by observers worldwide and used 
in this research. The optical spectroscopic data from ARAS site (\textit{Astronomical Ring for Amateur 
Spectroscopy}) was helpful in our analysis. We are most grateful to Paul Kuin for exploring the availability of 
Swift UV spectra contemporaneous with the near-IR  observations. GS acknowledges a 
WOS-A grant from the Department of Science and Technology (SR/WOS-A/PM- 2/2021). 
KLP acknowledges funding from the UK Space Agency. We thank John Rayner, Director 
IRTF, for Director’s Discretionary Time (2023B988) who made scheduling of this program 
possible and the IRTF telescope operator for assisting CEW with the observations. The 
Infrared Telescope Facility, is operated by the University of Hawaii under contract 
80HQTR19D0030 with the National Aeronautics and Space Administration. 
SS acknowledges partial support from a NASA Emerging Worlds grant to ASU 
(80NSSC22K0361) as well as support from his ASU Regents' Professorship. The x-ray 
data underlying this paper are available in the Swift archive at \url{https://www.swift.ac.uk/swift_live/} and
the HEASARC Browse archive at \url{https://heasarc.gsfc.nasa.gov/cgi-bin/W3Browse/w3browse.pl}. 
This work has made use of data from the European Space Agency (ESA) mission
{\it Gaia} (\url{https://www.cosmos.esa.int/gaia}), processed by the {\it Gaia}
Data Processing and Analysis Consortium (DPAC,
\url{https://www.cosmos.esa.int/web/gaia/dpac/consortium}). Funding for the DPAC
has been provided by national institutions, in particular the institutions
participating in the {\it Gaia} Multilateral Agreement.



\facilities{AAVSO, IRTF (SpeX), Swift, Gaia, DECaps}

\software{Astro Data Lab \citep{2020A&C....3300411N},  Astropy \citep{2018AJ....156..123A}, 
CLOUDY \citep{1998PASP..110..761F, 2023RMxAA..59..327C},
Spextool \citep{2004PASP..116..362C}, HEASOFT/XSPEC \citep{1996ASPC..101...17A}. }

\clearpage
\bibliographystyle{aasjournal}
\bibliography{v1716sco_refs_v2}{}

\end{document}